\documentstyle[epsfig,12pt]{article}
\begin{document}
%
%
%
%
\newcommand{\x}{\cdot}
\newcommand{\ra}{\rightarrow}
\newcommand{\pom}{\mbox{$\rm{\cal P}$omeron}}
\newcommand{\regge}{\mbox{$\rm{\cal R}$eggeon}}
\newcommand{\flux}{\mbox{$ F_{{\cal P}/p}(t, \xi)$}}
\newcommand{\ap}{\mbox{$\bar{p}$}}
\newcommand{\pp}{\mbox{$p$}}
\newcommand{\pap}{\mbox{$p \bar{p}$}}
\newcommand{\SPS}{\mbox{S\pap S}}
\newcommand{\xp}{\mbox{$x_{p}$}}
\newcommand{\et}{\mbox{${E_T}$}}
\newcommand{\etj}{\mbox{$\et ^{jet}$}}
\newcommand{\sumet}{\mbox{$\Sigma \et$}}
\newcommand{\sumetj}{\mbox{$\Sigma \et ^{jet}$}}
\newcommand{\pt}{\mbox{${P_t}$}}
\newcommand{\mpr}{\mbox{${ m_p}$}}
\newcommand{\mpi}{\mbox{${ m_\pi}$}}
\newcommand{\rs}{\mbox{$ \sqrt{s}$}}
\newcommand{\rsp}{\mbox{$ \sqrt{s'}$}}
\newcommand{\rsps}{\mbox{$ \sqrt{s} = 630 $ GeV}}
\newcommand{\lum}{\mbox{$\int {\cal L} {\rm dt}$}}
\newcommand{\T}{\mbox{$t$}}
\newcommand{\abt}{\mbox{${ |t|}$}}
\newcommand{\di}{\mbox{d}}
\newcommand{\sigdifjets}{\mbox{$ \sigma_{sd}^{jets}$}}
\newcommand{\sigpomjets}{\mbox{$ \sigma_{{\cal P}p}^{jets}$}}
\newcommand{\sigdiftot}{\mbox{$ \sigma_{sd}^{total}$}}
\newcommand{\dsig}   {\mbox{$ {{ d^2 \sigma        }\over{d \xi dt}} $}}
\newcommand{\dsigdif}{\mbox{$ {{ d^2 \sigma _{sd}  }\over{d \xi dt}} $}}
\newcommand{\sigpompom}{\mbox{$\sigma_{{\cal P}{\cal P}}^{tot}$}}
\newcommand{\sigdpe}   {\mbox{$\sigma_{DPE}$}}
\newcommand{\sigpomtot}{\mbox{$ \sigma_{{\cal P}p}^{tot}$}}
\newcommand{\sigpptot}{\mbox{$ \sigma_{pp}^{tot}$}}
\newcommand{\alamb}{\mbox{$\overline{\Lambda^{\circ}}$}}
\newcommand{\lamb}{\mbox{$\Lambda^{\circ}$}} 
\newcommand{\PRET}{\mbox{\Proton-\sumet}}
\begin{titlepage}
\begin{center}
{\large   }
{\large EUROPEAN ORGANIZATION FOR NUCLEAR RESEARCH}
\end{center}
\vspace{1 ex}
\begin{flushright}
{
1 July, 2002\\
}
\end{flushright}
\vspace{1 ex}
\begin{center}
{
\LARGE\bf
\rule{0mm}{7mm}{\boldmath A Study of Inclusive}\\ 
\rule{0mm}{7mm}{\boldmath Double--\pom --Exchange}\\
\rule{0mm}{7mm}{\boldmath in $\pap \ra p X \ap$ at \rs\ = 630 GeV}\\
}

\vspace{4ex}

A. Brandt$^{1}$, S. Erhan$^{a}$, A. Kuzucu$^{2}$, M. Medinnis$^{3}$,\\
N. Ozdes$^{2,4}$, P.E. Schlein$^{b}$, M.T. Zeyrek$^{5}$, J.G. Zweizig$^{6}$\\
University of California$^{*}$, Los Angeles, California 90024, U.S.A. \\
\vspace{2 ex}
J.B. Cheze, J. Zsembery \\
Centre d'Etudes Nucleaires-Saclay, 91191 Gif-sur-Yvette, France.
\end{center}
\vspace{2 ex}

\centerline{(UA8 Collaboration)}

\vspace{2 ex}

\begin{abstract} 
We report measurements of the inclusive reaction, $p \ap \ra p X \ap$, 
in events where either or both the beam--like final--state baryons were
detected in Roman-pot spectrometers and the central system was detected in
the UA2 calorimeter.
A Double-\pom -Exchange (DPE) analysis of these data and 
single diffractive data from the same experiment demonstrates that,
for central masses of a few GeV, the extracted \pom --\pom\ total cross 
section, \sigpompom , exhibits an enhancement which 
exceeds factorization expectations by an 
order-of-magnitude. This may be a signature for glueball production.
The enhancement is shown to be independent of uncertainties connected with
possible non--universality of the \pom\ flux factor.
Based on our analysis, we present DPE cross section predictions,
for unit (1 mb) \pom -\pom\ total cross section,
at the Tevatron, LHC and the 920 GeV fixed-target experiment, HERA-B.
\end{abstract}

\begin{center}
In press: European Physics Journal C \\
\end{center}
\vspace{3 ex}
\rule[.5ex]{16cm}{.02cm}
$^{*}$ Supported by U.S. National Science Foundation
Grant PHY-9986703. \\
$^{a}$ email:  samim.erhan@cern.ch \\
$^{b}$ email:  peter.schlein@cern.ch \\
$^{1}$ Present address: University of Texas, Arlington, U.S.A. \\
$^{2}$ Visitor from Cukurova University, Adana, Turkey; also supported by 
ICSC - World Lab.\\
$^{3}$ Present address: DESY, Hamburg, Germany \\
$^{4}$ Present address: Muscat Technical Industrial College (MTIC), 
Muscat/Oman \\
$^{5}$ Visitor from Middle East Tech. Univ., Ankara, Turkey; supported by 
Tubitak. \\
$^{6}$ Present address: California Institute of Technology, Pasadena, CA, 
U.S.A. \\
\end{titlepage}      

%
\setlength{\oddsidemargin}{0 cm}
\setlength{\evensidemargin}{0 cm}
\setlength{\topmargin}{0.5 cm}
\setlength{\textheight}{22 cm}
\setlength{\textwidth}{16 cm}
\setcounter{totalnumber}{20}
\clearpage\mbox{}\clearpage
\pagestyle{plain}
\setcounter{page}{1}
%

\section{Introduction}
\label{sect:intro}
\indent

\begin{figure}
\begin{center}
\mbox{\epsfig{file=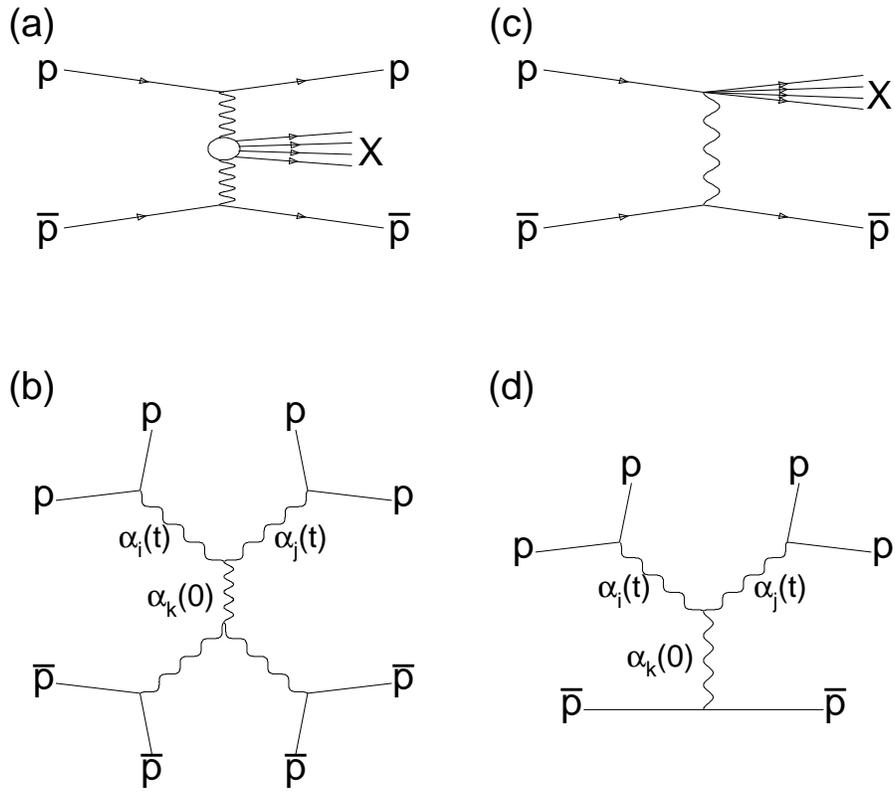,width=15cm}}
\end{center}
\caption[]{
(a) Inclusive double-\pom -exchange reaction 
(the central blob is the \pom -\pom\ interaction) and its corresponding
Triple-Regge diagram in (b); 
(c,d) Inclusive single diffractive reaction and its corresponding Triple-Regge 
diagram. In both cases, \pom -exchange dominance means $i = j = \pom$.
$k$ can be either \pom\ or \regge .
}
\label{fig:dpediag}
\end{figure}

We study the Double--\pom --Exchange (DPE) ``diffractive" process\cite{pompom}, 
depicted in Fig.~\ref{fig:dpediag}(a), 
using the reaction:
\vspace{2mm}
\begin{equation}
\bar{p} \, p  \,\,\,\, \rightarrow  \,\,\,\, \bar{p} \,\, X \,\, p.
\label{eq:pompom}
\end{equation}
%
When the final state $p$ and \ap\ both have large Feynman-$\xp$, 
the process proceeds with the exchange of two (virtual) gluon--rich 
colorless systems called \pom s. 
These systems, which carry a small fraction of the beam momentum of the two 
approaching hadrons, $ \Delta p/p = \xi = 1 - \xp$,
collide and constitute the entire effective interaction  
between the two beam particles.
This leads to the presence of ``rapidity--gaps", or regions of pseudorapidity
with no particles between the outgoing
$p$ and \ap\ and the central system, $X$.
The system $X$ with invariant mass, $M_X$, 
is the result of the \pom --\pom\ interaction.
To good approximation, $ {M_X}^2 = s' = \xi_1 \xi_2 s$
(we use the symbols, ${M_X}^2$ and $s'$, interchangably); thus, a given $M_X$
is produced at smaller $\xi$ values when the c.m. energy is larger.
Single diffractive processes (see Fig.~\ref{fig:dpediag}(c))
appear\cite{ua8dif} to be essentially pure \pom -exchange when $\xi < 0.03$.
We expect this to be also the case in DPE.
 
The DPE process is the closest we can come to pure gluon interactions.
As such, it may be a splendid glueball production process~\cite{glue}.
At the very high energies of the LHC, 
``diffractive hard scattering" in React.~\ref{eq:pompom} may have 
advantages as a relatively clean production mechanism of rare states. 
Diffractive hard scattering was proposed in Ref.~\cite{is} and discovered in 
$p\ap$ interactions by the UA8 experiment~\cite{ua8} 
at the \SPS --Collider (\rs\ = 630 GeV) and in $ep$ interactions 
by the H1 and ZEUS experiments at HERA\cite{herahard}
(see also Refs.~\cite{cdfhard,d0hard} for studies of these effects
at the Tevatron). 
First studies of hard scattering in React.~\ref{eq:pompom} were made at
the \SPS --Collider~\cite{ua1dpe} and at the Tevatron~\cite{cdfdpe,d0dpe}.
Based on UA8's observation of the ``super-hard" \pom\ in single
diffractive dijet production~\cite{ua8}, a small fraction ($\approx 10\%$) of  
all hard-scattering DPE events at the LHC may be high--mass gluon--gluon 
collisions.

In the present paper, we present final results on React.~\ref{eq:pompom}
from the UA8 experiment at the CERN \SPS --Collider.
UA8 was the first experiment in which data acquisition from
a large central detector was ``triggered" by  
the presence of outgoing beam-like protons or antiprotons.
The final-state baryons 
were measured in UA8 Roman--pot spectrometers~\cite{ua8spect}
which were installed in the outgoing arms of the same interaction region
as the UA2 experiment~\cite{ua2}; 
the central system, $X$, was measured in the UA2 calorimeter.
At 630 GeV center-of-mass energy, $M_X$ = 6.3 (18.9) GeV
when $\xi_1$ and $\xi_2$ are both equal to 0.01 (0.03).

Previous measurements~\cite{isr} of React.~\ref{eq:pompom} with
exclusive final states were made in $pp$ interactions at the CERN
Intersecting Storage Rings with c.m. energy, $\rs \, = 63$ GeV,
and with $\alpha$ beams~\cite{alpha} at $\rs \, = 126$ GeV.
The advantage of the present ten--times higher c.m. energy is that 
much smaller values of $\xi$ are accessible for a given produced $M_X$, 
thereby enhancing the purity of the \pom -exchange component.

As seen in Fig.~\ref{fig:dpediag}, Reaction~\ref{eq:pompom} is intimately 
related to the inclusive single--diffractive reactions:
%
\begin{equation}
\bar{p} \, p  \,\, \rightarrow  \, \, \bar{p} \, X  
\,\,\,\,\,\,\,\,\, {\rm or} \,\,\,\,\,\,\,\,\, 
\bar{p} \, p  \,\, \rightarrow  \, \,  X \, p 
\,\,\,\,\,\,\,\,\, , \,\,\,\,\,\,\,\,\, 
p \, p  \,\, \rightarrow  \, \,  X \, p. 
\label{eq:dif}
\end{equation}
%
In Reactions~\ref{eq:dif}, 
the \pom\ from one beam particle interacts with the 
second beam particle. 
Fig.~\ref{fig:dpediag}(c) depicts the single diffractive reaction, 
while Fig.~\ref{fig:dpediag}(b,d) show the 
corresponding (dominant) Triple-Regge diagrams of DPE and inclusive single 
diffraction.

At small four--momentum transfer\footnote{Throughout this paper, we refer
to the squared four-momentum transferred to the final--state p or \ap\ as the
``momentum transfer"}, $|t|$, 
triple--Regge predictions of inclusive single diffractive cross 
sections are found to increase more rapidly than do the observed cross 
sections~\cite{dino,es1}
and violate the unitarity bound above \rs\ = 2~TeV. 
The observed large damping effects in the data are believed to be due to 
multiple--\pom --exchange effects, which phenomenologically are equivalent
to a smaller effective \pom\ trajectory 
intercept with increasing energy~\cite{kpt,es2}. 
However, despite these unitarizing effects, effective vertex factorization 
appears to remain valid to an astonishing degree~\cite{es2}.
In the present analysis, we assume its validity. 
   
In terms of the Triple--Regge model, the cross section 
for React.~\ref{eq:dif} may be written as the product of the \pom -proton
total cross section, \sigpomtot , with the
flux factor for a \pom\ in the proton , \flux .
Since it is our working assumption that the same \flux\ describes the
\pom -proton vertices in both Reacts.~\ref{eq:dif} and \ref{eq:pompom},
the cross section for React.~\ref{eq:pompom}
is given by the product of the \pom -\pom\ total cross section, 
\sigpompom ($s'$), with two flux factors. 
See, however, the discussion in Sect.~\ref{sect:discuss}
on systematic uncertainties due to a possible non--universality of \flux .
The essential result of this paper will be shown to be insensitive to such 
effects.

The empirical \flux\ has been ``fine-tuned" in fits 
of the following equation to all available data on React.~\ref{eq:dif}
at the \SPS ~\cite{ua8dif} and ISR~\cite{albrowisr}:
\begin{equation}
{{d^2\sigma_{sd}} \over{d\xi dt}} \, = \, \flux \cdot \sigpomtot (s') \, = \,
[K \cdot |F_1(t)|^2 \cdot e^{bt} \cdot \xi^{1-2\alpha (t)}] \cdot
[\sigma_0 \cdot ((s')^{0.10} + R (s')^{-0.32})]. 
\label{eq:dsigdif}
\end{equation}
$|F_1(t)|^2$ is the Donnachie-Landshoff~\cite{dl1}
form factor\footnote{$F_1(t)={{4 m_p ^2 - 2.8t}
\over{4 m_p ^2 - t}}\, \x \, {1\over{(1-t/0.71)^2}}$}. 
The right-hand bracket in Eq.~\ref{eq:dsigdif}, the \pom -proton total cross 
section, 
is assumed  to have the same form that describes 
the $s$-dependence of real particle total cross sections.
The best values of the fitted parameters~\cite{ua8dif} in Eq.~\ref{eq:dsigdif} 
are\footnote{The fits are also consistent with the
existing CDF results at the Tevatron~\cite{es1,es2}}:
\begin{tabbing}
\hspace{5cm}\=$K\sigma _0$ \hspace{3mm}\= =\hspace{4mm}\=$0.72 \pm 0.10$ 
\hspace{5mm}\=mb GeV$^{-2}$\\
    \>$R$           \> =\> $4.0 \pm 0.6$\\
    \>$b$           \> =\> $1.08 \pm 0.20$   \>GeV$^{-2}$ \\
 \end{tabbing}
The effective \pom\ trajectory is found~\cite{es2} to be $s$--dependent
and, at the energy of the UA8 experiment (\rs\ = 630~GeV), is: 
\begin{equation}
\alpha(t) \, = \, 1 + \epsilon + \alpha' t + \alpha'' t^2 \, = \, 
1.035 + 0.165 t + 0.059 t^2
\label{eq:traj}
\end{equation}
while, over the ISR energy range ($s$ = 549 to 3840 GeV$^2$): 
\begin{tabbing}
\hspace{3cm}\=$\epsilon  (s)$ \hspace{6mm}\= =\hspace{4mm}\=
$(0.096 \pm 0.004) - (0.019 \pm 0.005) \cdot log (s/549)$.\\ 
\>$\alpha '  (s)$ \> = \> $(0.215 \pm 0.011) - (0.031 \pm 0.012) \cdot 
log (s/549)$.\\ 
\>$\alpha '' (s)$ \> = \> $(0.064 \pm 0.006) - (0.010 \pm 0.006) \cdot 
log (s/549)$.\\
\end{tabbing}

\begin{figure}
\begin{center}
\mbox{\epsfig{file=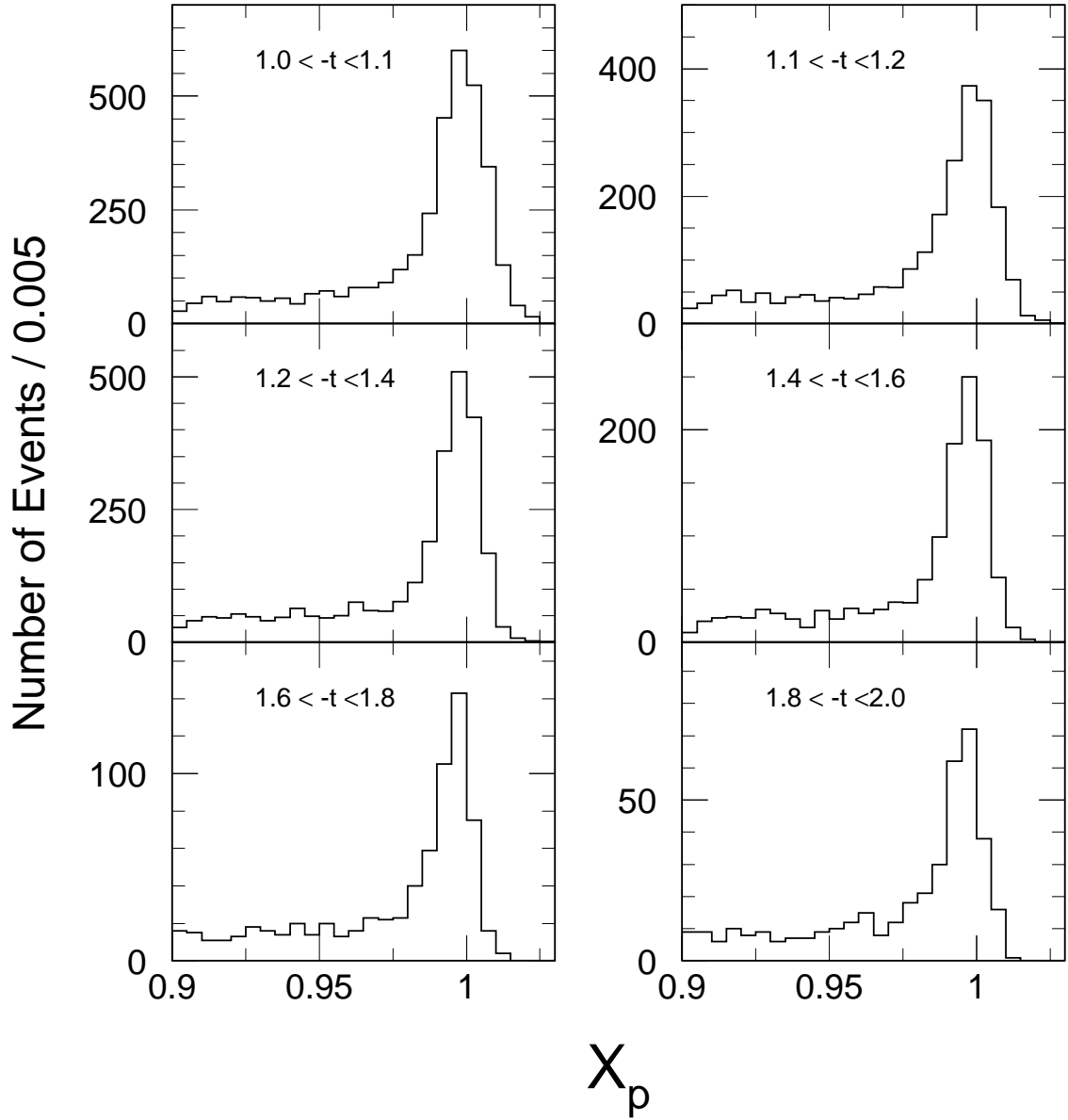,width=18cm}}
\end{center}
\caption[]{
Raw (uncorrected) 
Feynman-\xp\ distributions for different bins of momentum 
transfer (units are GeV$^2$)
in single-diffractive data from the UA8 experiment~\cite{ua8dif}.
As explained in the text, the \xp --dependence of the geometrical acceptance
is not responsible for the observed peaks. 
}
\label{fig:dndx}
\end{figure}

The quadratic term\cite{ua8dif} in $\alpha(t)$ corresponds to a 
``flattening"\footnote{This flattening is also claimed to be seen
by the ZEUS experiment~\cite{zeus} at DESY in photoproduction of
low-mass vector mesons ($\rho^0$ and $\phi^0$)}, 
or departure from linear behavior, 
of the effective \pom\ trajectory at high-$|t|$.
Direct evidence for this flattening of the trajectory can be 
obtained by looking at the behavior of the UA8 single diffractive
data~\cite{ua8dif} at large-$|t|$. 
Figure~\ref{fig:dndx} shows the observed Feynman-\xp\ distributions for 
different bands of $|t|$ between 1 and 2 GeV$^2$.
Since the geometrical acceptance\cite{ua8dif} 
depends linearly and weakly on \xp\ in this 
figure, the pronounced peaks near \xp\ = 1 reflect the physics of 
diffraction and are seen to persist up to $|t|$ of 2~GeV$^2$.
They are due to the (approximate) 
$1/M_X^2$ behavior of Triple-Regge phenomenology.
If the trajectory
did not flatten, but continued to drop linearly, the diffractive peak
would tend to disappear. For example, with a trajectory, 
$\alpha (t) = 1.08 + 0.25 t$, the peak would disappear at 
$-t = 2.3$~GeV$^2$ (corresponding to: $2\alpha (t) - 1 = 0$).
Thus, the persistence of the diffractive peak
in Fig.~\ref{fig:dndx} is the most direct evidence that
the effective \pom\ trajectory flattens at large-$|t|$.

The question arises as to whether  \pom --exchange is still dominant for 
$|t| > 1$~GeV$^2$, where most of the data in the present experiment are.
The self-consistency of our Triple--Regge analysis in describing 
both single--diffraction and double--\pom --exchange data is one supporting
argument. 
Another important point is that 
the set of all \SPS\ and ISR high--$|t|$ data agree~\cite{es1} with
a ``fixed--pole" description {\it without damping}.
Another argument is that the hard \pom\ structure found in the UA8 jet event 
analysis~\cite{ua8} is consistent 
with that found in the analysis of low-$|t|$ data at HERA\cite{h1}. 
Thus, our working assumption is that \pom --exchange dominates 
React.~\ref{eq:pompom} in the 
momentum--transfer range, $1 < -t < 2$~GeV$^2$.
Based on the results of earlier studies\cite{ua8dif} of diffraction, 
we can ignore Reggeon exchange when $\xi < 0.03$.

The differential cross section for the DPE process, React.~\ref{eq:pompom}, 
is:
\begin{equation}
{{d^6\sigma_{DPE}}\over{d\xi_1 d\xi_2 dt_1 dt_2 d\phi_1 d\phi2}} \, \, = \, \,
F_{{\cal P}/p}(t_1, \xi_1) \x F_{{\cal P}/p}(t_2, \xi_2) \x \sigpompom (s').
\label{eq:pompomdist}
\end{equation}
The variables, ($\xi_i, t_i, \phi_i$), describe each of the
emitted \pom s at the outer vertices in Fig.~\ref{fig:dpediag}(a), which
are uniquely given by the measurement of the associated outgoing 
$p$ (or \ap ) in the final state.
Although there is no explicit 
$\phi$--dependence on the right-hand-side of Eq.~\ref{eq:pompomdist} and the 
\pom s are emitted independently and isotropically, $\phi$ correlations do 
result, because significant regions in the 6-dimensional space, 
($\xi_1, \, t_1, \, \phi_1, \, \xi_2, \, t_2, \, \phi_2$), 
are unphysical and give $s' < 0$.
This point is discussed further in Sect.~\ref{sect:mc} in connection with 
Monte--Carlo generation of events according to Eq.~\ref{eq:pompomdist}.

\begin{figure}
\begin{center}
\mbox{\epsfig{file=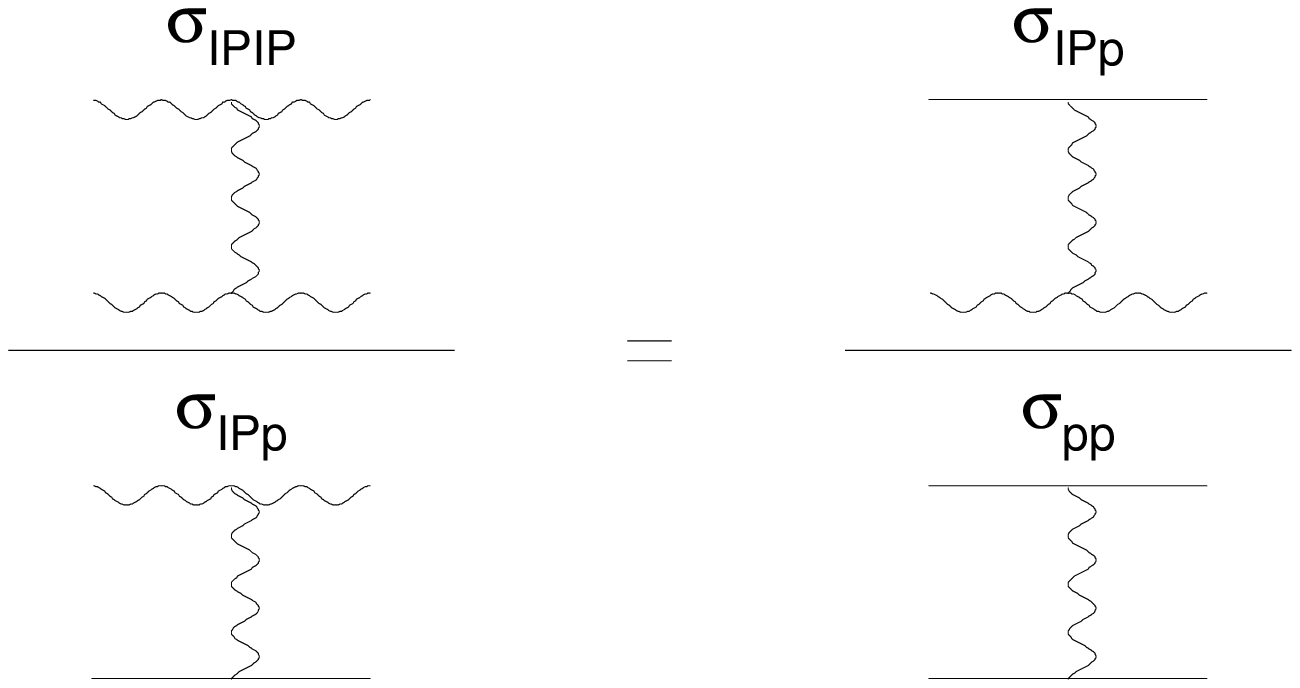,width=18cm}}
\end{center}
\caption[]{
Ratios of forward elastic amplitudes which are equal if factorization is 
valid. The optical theorem implies that the following relation between 
the corresponding total cross sections should be valid if \pom -exchange
dominates: $\sigpompom = (\sigpomtot )^2 / \sigpptot$.
}
\label{fig:factor}
\end{figure}

Using Eq.~\ref{eq:pompomdist}, our goal is to extract \sigpompom\ from our 
data on React.~\ref{eq:pompom} and to determine its energy ($s'$) dependence.
In particular, we wish to know whether there are enhancements at small $s'$
which could be due to a strong \pom -\pom\ interaction and possible
glueball production.
In the large--$s'$ region where
\pom -exchange dominates, \sigpompom\ is related by factorization to
the \pom -proton and proton-proton total cross sections:
\begin{equation}
\sigpompom (s') \, \, \, \, = \, \, \, \,
{  {[\sigpomtot(s')]^2}  \over  { \sigma _{pp}^{total}(s') }        }.
\label{eq:factorize}
\end{equation}
This is seen with reference to the ratios of forward elastic
amplitudes for the three processes shown in Fig.~\ref{fig:factor}.
A generalized optical theorem~\cite{almueller,others} then leads to 
Eq.~\ref{eq:factorize} between total cross sections. 
Ryskin~\cite{ryskin} has pointed out that
the three cross sections must be evaluated at the same value of $s'$.

Despite the fact that the cross sections, \sigpomtot\ and \sigpompom , 
can only be extracted from data in product with the constant $K$ in \flux\
($K^2$ in the case of \sigpompom\ and $K$ in the case of \sigpomtot;
see Eqs.~\ref{eq:dsigdif} and \ref{eq:pompomdist}), 
we see in Eq.~\ref{eq:factorize}
that such factors of $K$ cancel. 
{\it Thus, the factorization test does not require knowledge of $K$}. 
However, absolute values of either \sigpomtot\ or \sigpompom\
can only be given for an assumed value of $K$,
for example by using the Donnachie--Landshoff model~\cite{dl1} with 
$K =9 \beta^2/(4\pi^2) = 0.74$~GeV$^{-2}$, 
which arises from an analysis of elastic scattering data. 
Although different multi-\pom -exchange effects in diffraction and 
elastic scattering mean that this value of $K$ is only approximate, 
we nonetheless do quote values for \sigpompom\ in the closing sections 
of this paper, assuming $K = 0.74$~GeV$^{-2}$.

After describing the experiment and the event selection, 
we discuss the Monte--Carlo event generation of React.~\ref{eq:pompom},
and the determination of \sigpompom .
In the process, we compare the results from two different data samples, 
one in which {\it both} $p$ and \ap\ are detected and the other in which only
the $p$ {\it or} the \ap\ is detected.
We then test the factorization relation and come to
our conclusions which demonstrate, among other things, an overall
self-consistency of our phenomenological description of single diffraction
and double-\pom -exchange.  
Based on these conclusions, we calculate
predictions for double-pomeron-exchange yields at the Tevatron and LHC
and also at the HERA-B fixed--target experiment.

\section{Apparatus \& trigger}
\label{sect:apparatus}
\indent

Detailed descriptions of the UA8 apparatus, its properties, triggering
capabilities and interface to the UA2 experiment~\cite{ua2} are given 
elsewhere~\cite{ua8spect}. Thus we only provide a brief summary of
the spectrometer here.

\begin{figure}
\begin{center}
\mbox{\epsfig{file=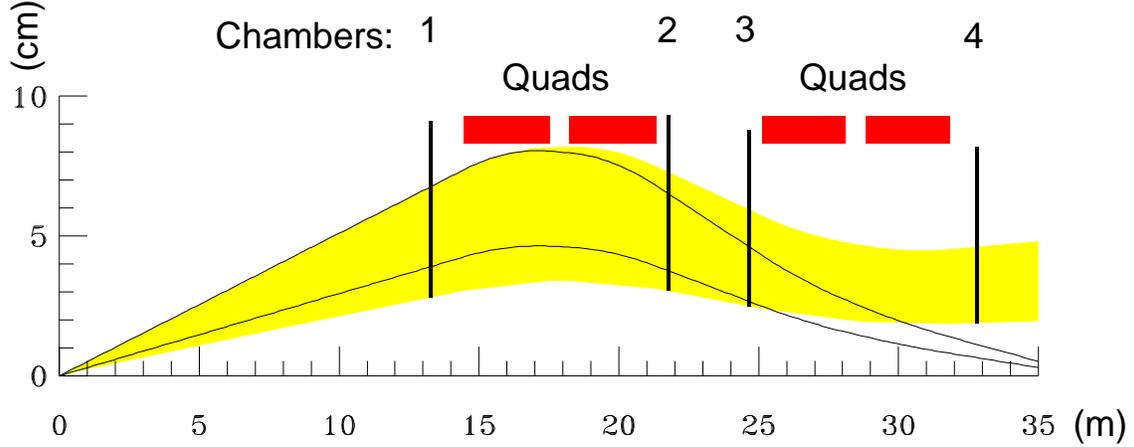,width=15cm}}
\end{center}
\caption[]{
Particle trajectories through a UA8 Roman-pot spectrometer.
The labels, ``Quads", refer to the low-$\beta$ machine quadrupole magnets.
The center of the UA2 detector is at $z = 0$ at the left of the sketch. 
The vertical lines indicate the positions
of the UA8 wire chambers in the Roman-pots.
The solid curves show the trajectories of 300 GeV particles ($\xp \sim 0.95$),
as described in the text.
The shaded area shows the allowed trajectories for
$\xp = 1$ tracks.
}
\label{fig:ua8traj}
\end{figure}

UA8 constructed Roman-pot spectrometers~\cite{ua8spect} 
in the same interaction region as UA2,
in order to  measure the outgoing ``beam-like" $p$ and/or \ap\ in 
React.~\ref{eq:pompom} or React.~\ref{eq:dif}, together with the central 
system using the UA2 calorimeter system~\cite{ua2}.
There were 
four Roman-pot spectrometers (above and below the beam pipe in each arm)
which measured $p$ and/or \ap\ with $\xp > 0.9$ and $0.8 < |t| < 2.5$ GeV$^2$. 
Fig.~\ref{fig:ua8traj} shows one spectrometer with
the trajectories of 300 GeV particles ($\xp \sim 0.95$)
emerging from the center of the intersection region with minimum and maximum 
acceptable angles (solid curves). 
The lower (upper) edge of the shaded area corresponds to the
minimum (maximum) accepted angles of $\xp = 1$ tracks.
The trajectory corresponding to the lower edge of the shaded region is 
12 beam widths ($\sigma$) from the center of the circulating beam orbit.

Particle momenta in the Roman pot spectrometers were calculated in real--time by
a dedicated special--purpose processor system~\cite{ua8spect,jgz}, thereby
providing efficient low--rate $p$ and \ap\ triggers.
Improved final-state proton and/or antiproton momenta
are calculated offline 
using the reconstructed vertex position (if it exists),
given by the UA2 central chamber system~\cite{ua2}, 
and points reconstructed from hits in Roman pot chambers
1, 2 and 3. Chamber 4 was also used in the fit, if a track traversed it.

\begin{figure}
\begin{center}
\mbox{\epsfig{file=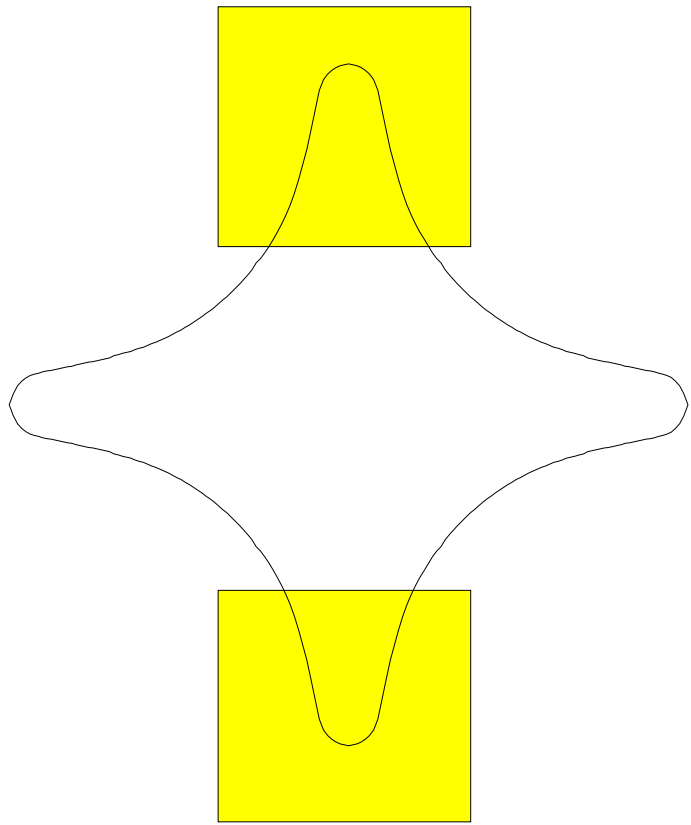,width=14cm}}
\end{center}
\caption[]{
UA8 spectrometer aperture viewed from the interaction region. The
shaded rectangles indicate the sensitive regions of the first wire chambers at a
distance $z = 13$~m from the interaction region center.  The curved line
indicates the walls of the beam pipe inside the quadrupole magnets.
}
\label{fig:aperture}
\end{figure}

Figure~\ref{fig:aperture} shows a ``beams-eye'' view of the UA8 chamber
aperture which is closest to the center of the interaction region.
The four-lobed curve in the figure illustrates the contour of the 
beam pipe which matches that of the quadrupole-magnet pole pieces.
The overlap between the beam pipe and rectangular chambers above and below
the beam illustrates the limited azimuthal ranges in which a
particle may be detected. 
These are centered at $\phi \sim 90^{\circ}$ and $\phi \sim 270^{\circ}$.
Data were recorded with the bottom edge of each pot set, in different runs,
at either 12 beam widths (12$\sigma$) or 14$\sigma$ 
from the beam axis.

\begin{figure}
\begin{center}
\mbox{\epsfig{file=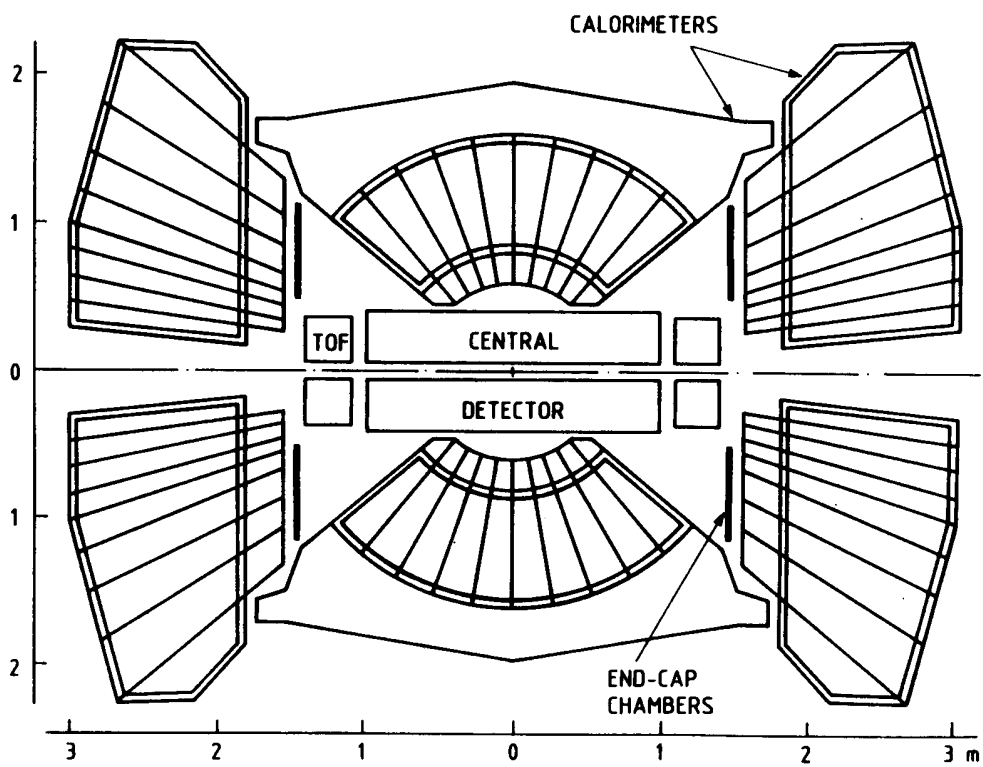,width=15cm}}
\end{center}
\caption[]{
A cross sectional view of the upgraded UA2 apparatus. 
Detectors which were used for the measurements reported here 
are the Calorimeters, the Time--Of--Flight (TOF) counters
and the Silicon Vertex Detector within the Central Detector assembly.
The TOF counters covered pseudorapidity from 2.3 to 4.1 in each arm.
}
\label{fig:ua2}
\end{figure}

The upgraded UA2 calorimeter system~\cite{ua2}, shown in Fig.~\ref{fig:ua2}, 
covered the polar angular range, 
$6^{\circ} < \theta < 174^{\circ}$, and was used to 
measure the central system, $X$.
In order to isolate React.~\ref{eq:pompom} from other (background) events,
rapidity--gaps are imposed {\it offline} between $p$ and $X$, and between \ap\ 
and $X$, by requiring the absence of charged--particle hits in 
the UA2 Time--Of--Flight (TOF) counters.
These counters are indicated in Fig.~\ref{fig:ua2} and cover the 
range of pseudorapidity, $2.3 < |\eta| < 4.1$, in both 
arms ($2^{\circ}$--$12^{\circ}$ and 
$168^{\circ}$--$178^{\circ}$). Since the TOF counters have some overlap 
with the small--angle region of the end--cap calorimeters, the calorimeter 
minimum acceptance angle for the events considered here is increased from
$6^{\circ}$ to $12^{\circ}$ in both arms. 

\subsection{Triggering}
\label{sect:trigger}
\indent

Since the main goal of the UA8 experiment was to make
measurements of hard-diffraction scattering
in React.~\ref{eq:dif}~\cite{is,ua8}, 
UA8 was interfaced to the UA2 data acquisition system, 
which allowed the
formation of triggers based on various combinations of $p$ and/or \ap\ 
momenta and transverse energy in the UA2 calorimeter system. 
Parallel triggers were
also employed to yield samples of elastic and inelastic diffraction reactions
with no conditions on the energy in the calorimeter system.

\begin{figure}
\begin{center}
\mbox{\epsfig{file=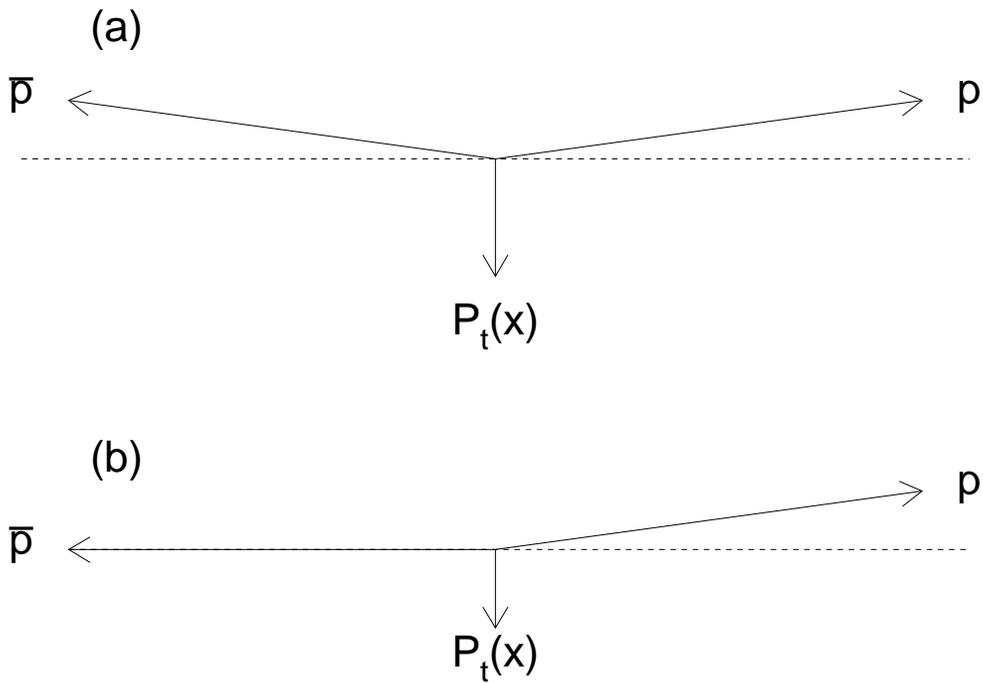,width=14cm}}
\end{center}
\caption[]{
(a) Side view sketch of an accepted event in which both $p$ and \ap\
go into the UP spectrometers. 
In this case, the central system recoils downward with
a minimum \pt\ of $\sim 2$~GeV.
(b) Sketch of an event triggered on $p$ {\it or} \ap . In this case,
the unobserved $p$ (or \ap ) has momentum transfer close to zero on average
and the central system recoils downward with a minimum \pt\ of $\sim 1$~GeV.
}
\label{fig:dpevector}
\end{figure}

In order to find evidence for React.~\ref{eq:pompom}, one of the supplementary
triggers required detection of a non-collinear $p$ and \ap\ pair.
The $p$ and \ap\ were both required to be either in the ``UP" spectrometers
(above the beam pipe), as shown in Fig.~\ref{fig:dpevector}(a),
or in the ``DOWN" spectrometers (below the beam pipe).

During the 1989 run, 1297 events were recorded in which both $p$ and \ap\ 
tracks were detected and the calorimeter system had a total recorded energy  
greater than 0.25 GeV. 
The remainder of the event-selection procedure for these events
is described in Sect.~\ref{sect:and}.

The essential topology characteristic of these events is summarized
in Table~\ref{tab:sigmas}. 
It is seen that, when both $p$ and \ap\ have $\xp > 0.95$, 48\% of the events 
have rapidity--gaps in both arms (with pseudorapidity, $2.3 < |\eta| < 4.1$). 
However, when one or the other of the tracks has $\xp < 0.95$, the percentage 
which possess rapidity--gaps in both arms falls to only a few percent. 
Thus, the first class of events appears to constitute a unique set, distinct 
from those events in which either $p$ or \ap\ have $\xp < 0.95$.

A secondary data sample of React.~\ref{eq:pompom} was extracted from
data which were triggered by requiring that {\it either} $p$ {\it or} \ap\
is observed, as shown in Fig.~\ref{fig:dpevector}(b). 
In these events, since there is no selection bias on momentum transfer, $t$, 
of the undetected particle, its natural distribution prevails,
with an average value, $|t| \approx 0.2$~GeV$^2$.
There were 62,627 such events recorded, after offline cuts for pile-up, 
beam-pipe geometry and halo cuts are made~\cite{ua8dif}.
These data are dominated by single diffraction, React.~\ref{eq:dif}. 
However, we show in Sect.~\ref{sect:or} that the offline imposition of the 
rapidity--gap veto in {\it both} arms isolates 
React.~\ref{eq:pompom} in this data sample.

\section{Event selection}
\label{sect:evsel}
\indent

We henceforth refer to the event sample for which both $p$ {\it and} \ap\ were
required in the trigger as the ``AND" data sample. The events for
which either $p$ {\it or} \ap\ are detected are referred to as 
the ``OR" data sample.

\subsection{``AND" data sample}
\label{sect:and}
\indent

The four constraints of energy-momentum conservation
can be examined in individual events 
because the entire final state of React.~\ref{eq:pompom} is seen in the
Roman-pot spectrometers and the calorimeter system.

\begin{figure}
\begin{center}
\mbox{\epsfig{file=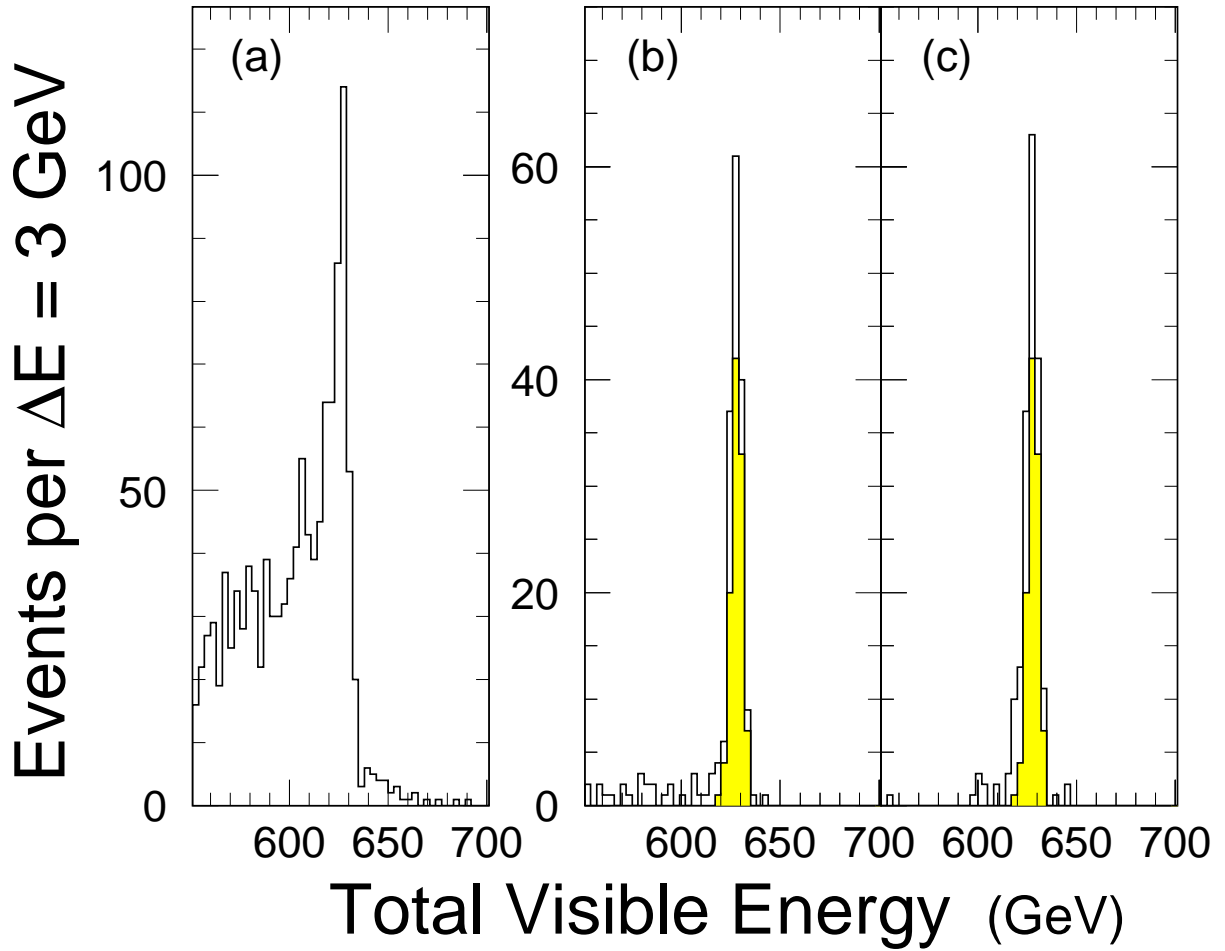,width=17cm}}
\end{center}
\caption[]{
(a) Total visible energy in ``AND" triggered events (1297)  
with total calorimeter energy, $\Sigma E > 250$~MeV, selected offline;
(b) Open histogram contains events (188) after veto using TOF counters,
as discussed in the text. 
Shaded events (107) are after momentum conservation cuts 
(Figs.~\protect\ref{fig:pcuts});
(c) Open histogram are events (193) 
after momentum conservation cuts, but before TOF cuts. 
Shaded events are after the TOF cuts (107 events).
}
\label{fig:etot}
\end{figure}

Fig.~\ref{fig:etot}(a) shows the distribution of total visible energy
($p$ and \ap\ and central system, $X$) for these events.
Although there is clearly a component of events which possess the full available
energy of 630 GeV, a significant fraction of the events have less energy.
Fig.~\ref{fig:etot}(b) shows the same distribution, but for those 
events in which rapidity--gaps have been imposed using 
the Time--Of--Flight (TOF) counters in both arms.
188 events remain in a clean signal at 630 GeV.

\begin{figure}
\begin{center}
\mbox{\epsfig{file=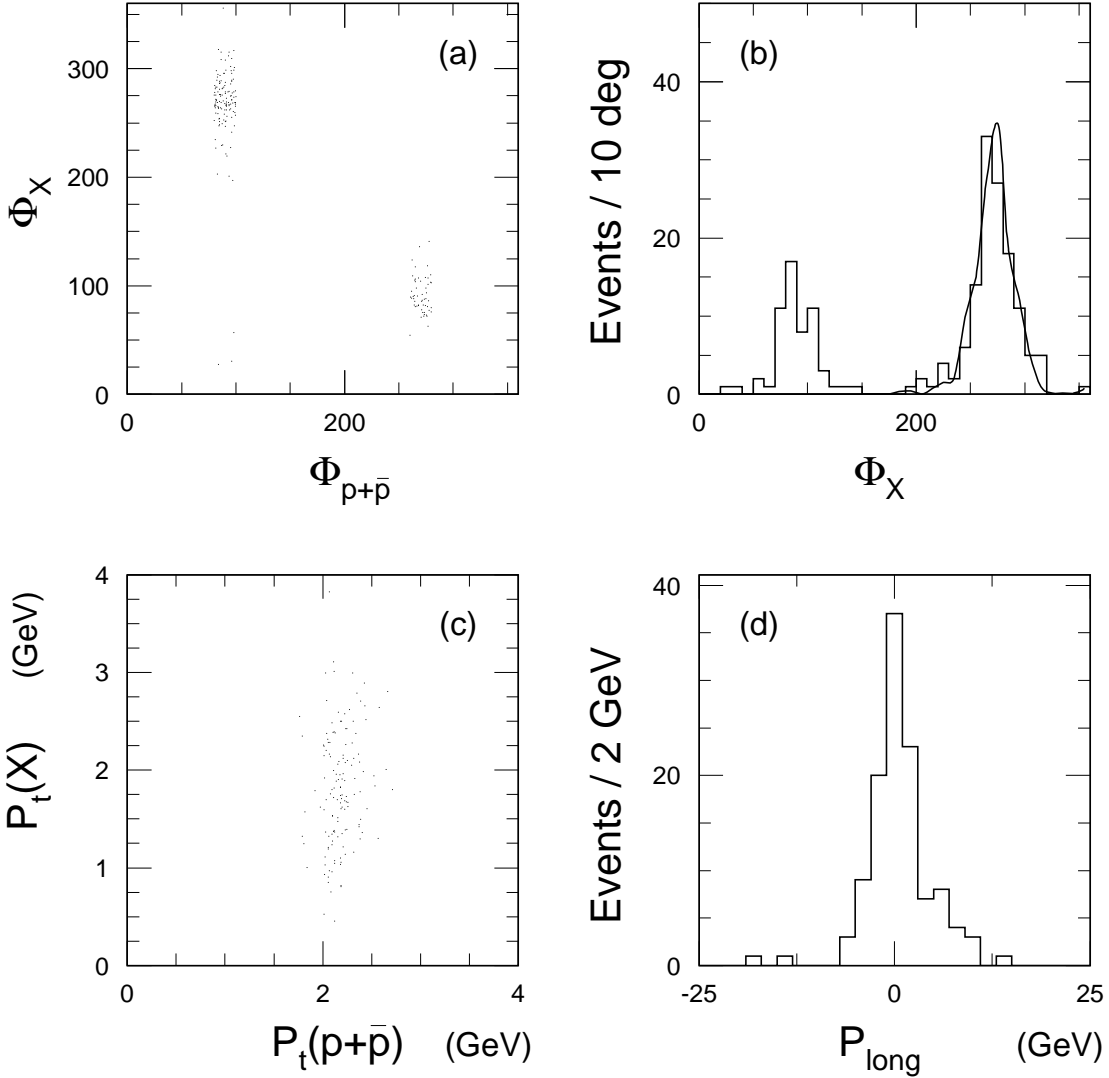,width=16cm}}
\end{center}
\caption[]{
``AND" events (188) after TOF veto selection (see text);
(a) Scatter plot of the azimuthal angles of the $p$ + \ap\ system 
and the summed calorimeter momentum vector; 
(b) $\Phi_X$ projection of (a). The curve on the right-hand peak is a 
Monte--Carlo simulation described in the text;
(c) $|\vec{P}_t(p + \ap)|$ vs. $\vec{P}_t(X)$ measured in calorimeter, 
for events (139) which satisfy $ \Phi_X$ selection in the bands, 
$90^{\circ} \, \pm \, 20^{\circ}$ and $270^{\circ} \, \pm \, 20^{\circ}$.
(d) Summed longitudinal momentum, $\Sigma P_{long}$, 
of $p$, \ap , and calorimeter for events (126) 
which satisfy the selection, $1 < \vec{P}_t(X) < 3$~GeV.
107 events satisfy the cut, $|\Sigma P_{long}| < 7$~GeV.
}
\label{fig:pcuts}
\end{figure}

Having seen that the energy constraint is well satisfied,
we now consider the three momentum constraints. 
As implied by Fig.~\ref{fig:dpevector}(a),
a minimum accepted transverse momentum of $\sim 1$ GeV for each of
$p$ and \ap\ corresponds to 
a net transverse momentum imbalance, $\pt > 2$ GeV, which
is compensated for by a corresponding (opposite) momentum vector in the UA2
calorimeter system. 
In order to observe this, we
define a summed momentum vector in the calorimeter.
The cell energies
observed in the UA2 calorimeter system are summed up as (massless)
vectors to approximate the total momentum vector, $\vec{P}(X)$, of the
system $X$ in React.~\ref{eq:pompom}.
The azimuthal angle of $\vec{P}(X)$, $\Phi_X$, is plotted
in Fig.~\ref{fig:pcuts}(a) vs. the azimuthal angle of the summed momentum
vector of the final-state $p$ and \ap\ particles.
There are peaks seen at $90^{\circ}$ and
$270^{\circ}$, corresponding to the cases where both $p$ and \ap\ are both in
their DOWN spectrometers or both in their UP spectrometers (as sketched in
Fig.~\ref{fig:dpediag}(b)), respectively.
Although $\Phi_X$ has no acceptance or trigger bias, in both cases it is
seen to be opposite the azimuthal angle of the summed \pap\ momentum vector
in the figure.

The projection of the points in Fig.~\ref{fig:pcuts}(a) on the $\Phi_X$
axis is shown in Fig.~\ref{fig:pcuts}(b).
The different intensities in the two peaks are due to small differences 
in the distances of the 
Roman-pots from the beam axis in the two spectrometers, 
resulting in a mismatch in their
low--\pt\ cutoffs.
The solid curve is a Monte--Carlo calculation and shows that the width of
the peaks is understood.
We select 139 events with $\Phi_X$ in the bands
$90^{\circ} \, \pm \, 20^{\circ}$ and $270^{\circ} \, \pm \, 20^{\circ}$.

Although the summed transverse momentum of $p$ and \ap , $\pt(\pp + \ap)$, 
drops off sharply below 2 GeV,
the transverse component of the calorimeter vector, $P_t(X)$,
shown in Fig.~\ref{fig:pcuts}(c)
displays a much broader distribution 
due to the resolution of the calorimeter and the fact that,
at small particle energies, some energy is lost before the particles reach 
the sensitive volume of the device. 
Fig.~\ref{fig:ptcomp}(a) shows the transverse projection of the
calorimeter momentum vector together with the result
of a Monte--Carlo simulation~\cite{ua2sim,thesis} 
of the UA2 calorimeter system.
This shows that the UA2 calorimeter simulation software does a good
job in describing the calorimeter's low energy response.
We select 126 events with $1 < \pt(X) < 3$ GeV for further analysis.

The degree of longitudinal momentum balance is demonstrated in 
Fig.~\ref{fig:pcuts}(d), a histogram of total longitudinal momentum, 
$\Sigma P_{long}$, which includes the $p$, \ap\ and calorimeter 
longitudinal energies.
A final sample of 107 DPE events satisfies the selection,
$|\Sigma P_{long}| < 7$ GeV.
The shaded histogram in Fig.~\ref{fig:etot}(b) shows the total
visible energy for these events and demonstrates how well the energy constraint
is satisfied.
Fig.~\ref{fig:etot}(c) is essentially the same as Fig.~\ref{fig:etot}(b),
except that the order of the TOF veto and momentum--conservation
cuts is inverted.

Table~\ref{tab:mc} summarizes the event losses due to the cuts described here.
They are compared with the effect of the same cuts on the Monte--Carlo
generated events discussed in Sect.~\ref{sect:mc}.
The similarity between the two sets of numbers implies that most of the
188 events shown in Figs.~\ref{fig:etot}(b) and~\ref{fig:pcuts}(a,b) 
are in fact real examples of React~\ref{eq:pompom}.

An additional point can be made that there is an insignificant contribution
in the data sample from events in which the observed proton comes from a
diffractively--produced low--mass system (Baksay et al. \cite{baksay}
measured that this occurs ($12 \pm 2.5$)\% of the time). 
Such events would lead to an asymmetry and tail on the low side of the total 
visible energy distribution in Fig.~\ref{fig:etot}. 
Although a small tail of this type does exist, it 
disappears when the momentum conservation cut is made.
Thus we conclude that the rapidity--gap veto combined with
momentum conservation eliminates this source of background.

\subsection{``OR" data sample}
\label{sect:or}
\indent

As remarked above, the ``OR" triggered sample is dominated by
React.~\ref{eq:dif}. 
However, the small component which is React.~\ref{eq:pompom}
can be isolated by selecting those events which possess
rapidity--gaps in both arms.

\begin{figure}
\begin{center}
\mbox{\epsfig{file=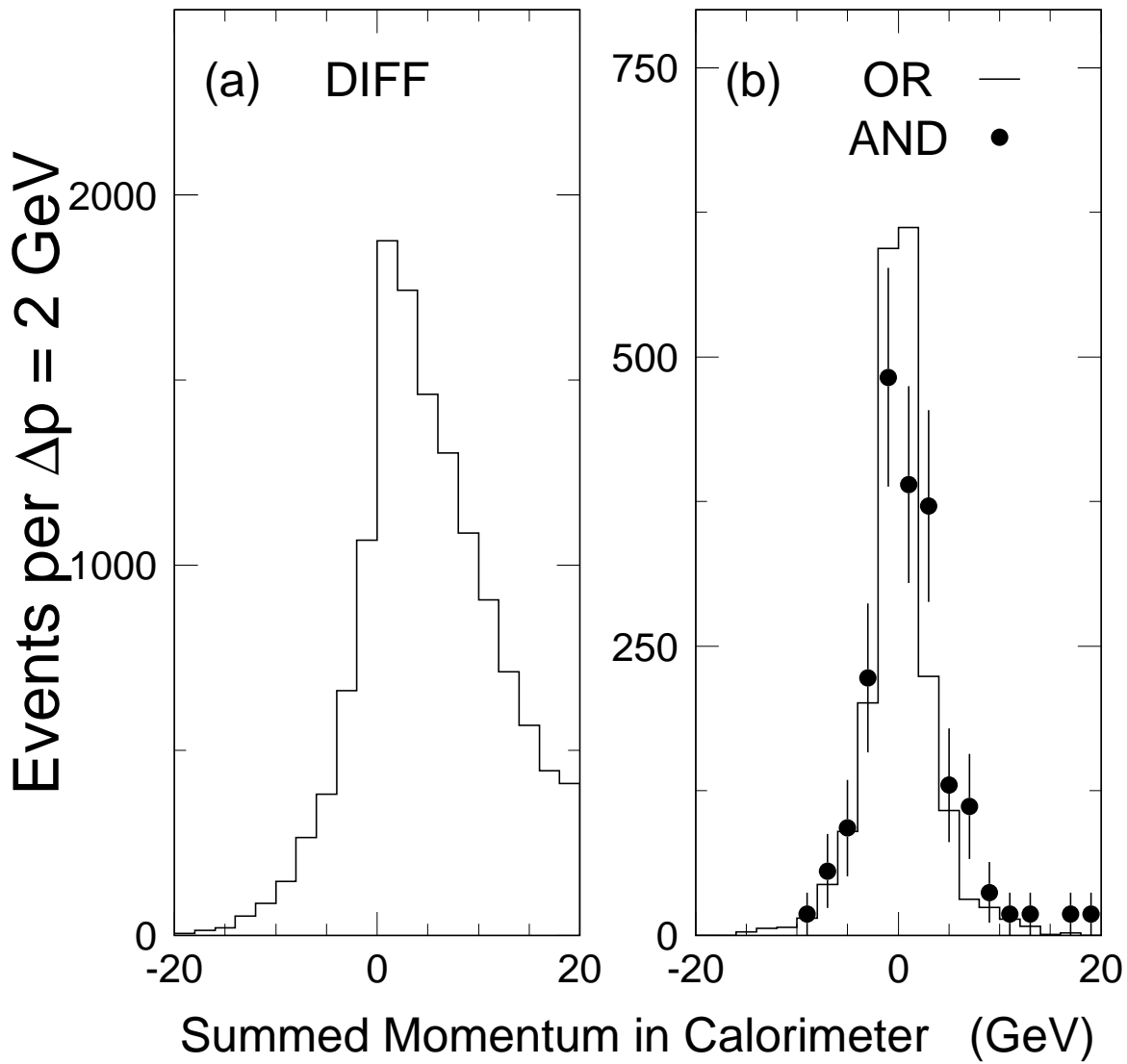,width=16cm}}
\end{center}
\caption[]{
For ``OR" triggered events, comparison of the summed longitudinal momentum in 
the calorimeter, with and without rapidity--gap veto (TOF).
(a) Without TOF veto (the single--diffractive data sample); 
events are plotted
on the positive axis if their summed momentum is in the hemisphere opposite
the observed trigger particle. 
Events are required to have at least 250 MeV energy in
the UA2 calorimeter system (partial sample, 15,080 events);
(b) With rapidity--gap veto (1985 events). 
Solid points show the 107 ``AND" events, normalized to the ``OR" data.
}
\label{fig:orplcomp}
\end{figure}

A signature which distinguishes React.~\ref{eq:pompom}
from React.~\ref{eq:dif} is the presence of a longitudinally
forward-backward symmetric distribution of particles in the UA2 calorimeter.  
Fig.~\ref{fig:orplcomp}(a) shows distribution of the summed longitudinal
momentum component of all struck cells in the calorimeter for 
the triggered ``OR" data sample.
In constructing this plot, each event is
plotted on the negative side if the summed vector is in the same hemisphere
as the observed trigger particle, and on the positive side if the vector
is in the opposite hemisphere. A large asymmetry is seen favoring
the hemisphere opposite the trigger particle, as expected for 
React.~\ref{eq:dif}. 

Fig.~\ref{fig:orplcomp}(b) shows the subset of events in
Fig.~\ref{fig:orplcomp}(a) which have no hits in any of the
UA2 Time--Of--Flight counters. 
This corresponds to rapidity--gaps in the range $2.3 < |\eta| < 4.1$
in both arms. 
The forward-backward asymmetry seen in the calorimeter system
disappears. 
The ``AND" events are also plotted in Fig.~\ref{fig:orplcomp}(b)
and are seen to have the same summed calorimeter momentum distribution 
as do the ``OR" events.
We take the events in the resulting symmetric distribution as 
the candidates for React.~\ref{eq:pompom}.

\begin{figure}
\begin{center}
\mbox{\epsfig{file=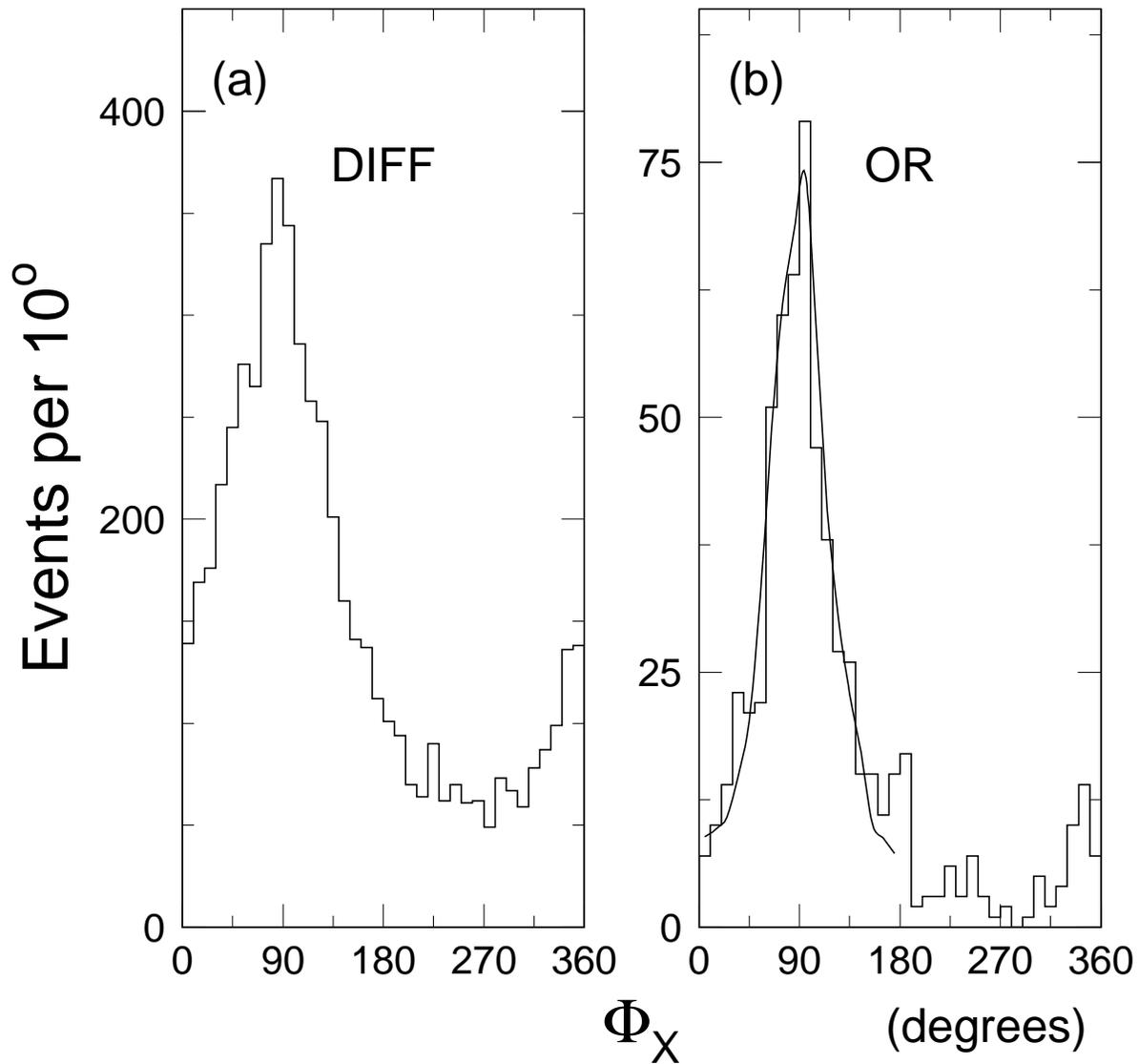,width=16cm}}
\end{center}
\caption[]{
Calorimeter azimuthal angle, $\Phi_X$.
(a) single-diffractive data sample, with $p$ or \ap\ in DOWN spectrometer
($\Phi_{\pp + \ap}$ selection in the band, $270^{\circ} \, \pm \, 20^{\circ}$) 
and with total calorimeter energy, $\Sigma E > 250$~MeV (5547 events);
(b) Same as (a), but after TOF veto selection to obtain (partial)
``OR" data sample (635 events). 
The curve in (b) is the Monte--Carlo simulation described in the text.
}
\label{fig:orphicomp}
\end{figure}

The next step in the selection of React.~\ref{eq:pompom} is to look
at the equivalent of Fig.~\ref{fig:pcuts}(b) for this sample,
namely the azimuthal angle, $\Phi_X$, of the summed calorimeter vector. 
Fig.~\ref{fig:orphicomp}(a) shows its distribution for those 
single-diffractive events in which the observed $p$ or \ap\ is seen in the 
DOWN spectrometer. Not surprisingly, a correlation is seen between the
azimuthal angles of the observed $p$ or \ap\ and the summed calorimeter 
vector.
However, when we make the ``OR" event selection, by imposing the rapidity--gap 
condition using the TOF counters, we see in Fig.~\ref{fig:orphicomp}(b)
that the correlation becomes much stronger. The distribution is broader 
than that seen in Fig.~\ref{fig:pcuts}(b) for the ``AND" events because
of the unknown (small) \pt\ of the unobserved final--state $p$ or \ap\ and 
also because of the smaller energy in the calorimeter.
The Monte--Carlo simulation of the UA2 calorimeter
is in reasonable agreement with the observed distribution.
We select 698 ``OR" events, with $\Phi_X$ either in the range
$90^{\circ} \, \pm \, 20^{\circ}$
or $270^{\circ} \, \pm \, 20^{\circ}$.

\begin{figure}
\begin{center}
\mbox{\epsfig{file=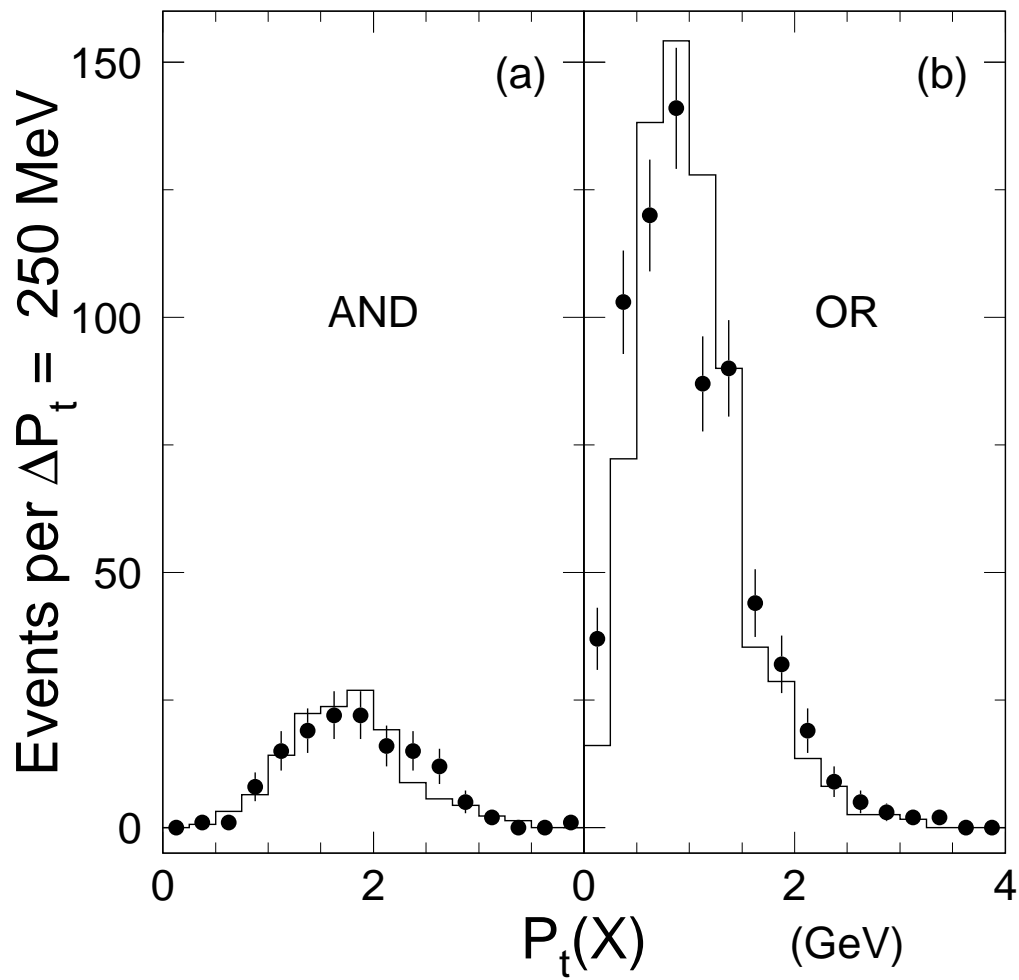,width=14cm}}
\end{center}
\caption[]{
Transverse momentum, \pt , measured in the UA2 calorimeter system:
(a) ``AND" data sample, projection of Fig.~\ref{fig:pcuts}(c) (139 events
);
(b) ``OR" data sample (698 events, after TOF and $\Phi_X$ cuts).
The histograms~\protect\cite{thesis} are from a Monte--Carlo 
simulation of the UA2 calorimeter response.
}
\label{fig:ptcomp}
\end{figure}

Finally, for these ``OR" candidates, 
we examine the summed transverse momentum in the calorimeter
shown in Fig.~\ref{fig:ptcomp}(b). Because this vector is opposite
only one observed vector of the $p$ or \ap , its average value
is less than that seen in Fig.~\ref{fig:ptcomp}(a) for the ``AND" sample.
However, it also is in reasonable agreement with the Monte--Carlo simulation
of the UA2 calorimeter.

\begin{figure}
\begin{center}
\mbox{\epsfig{file=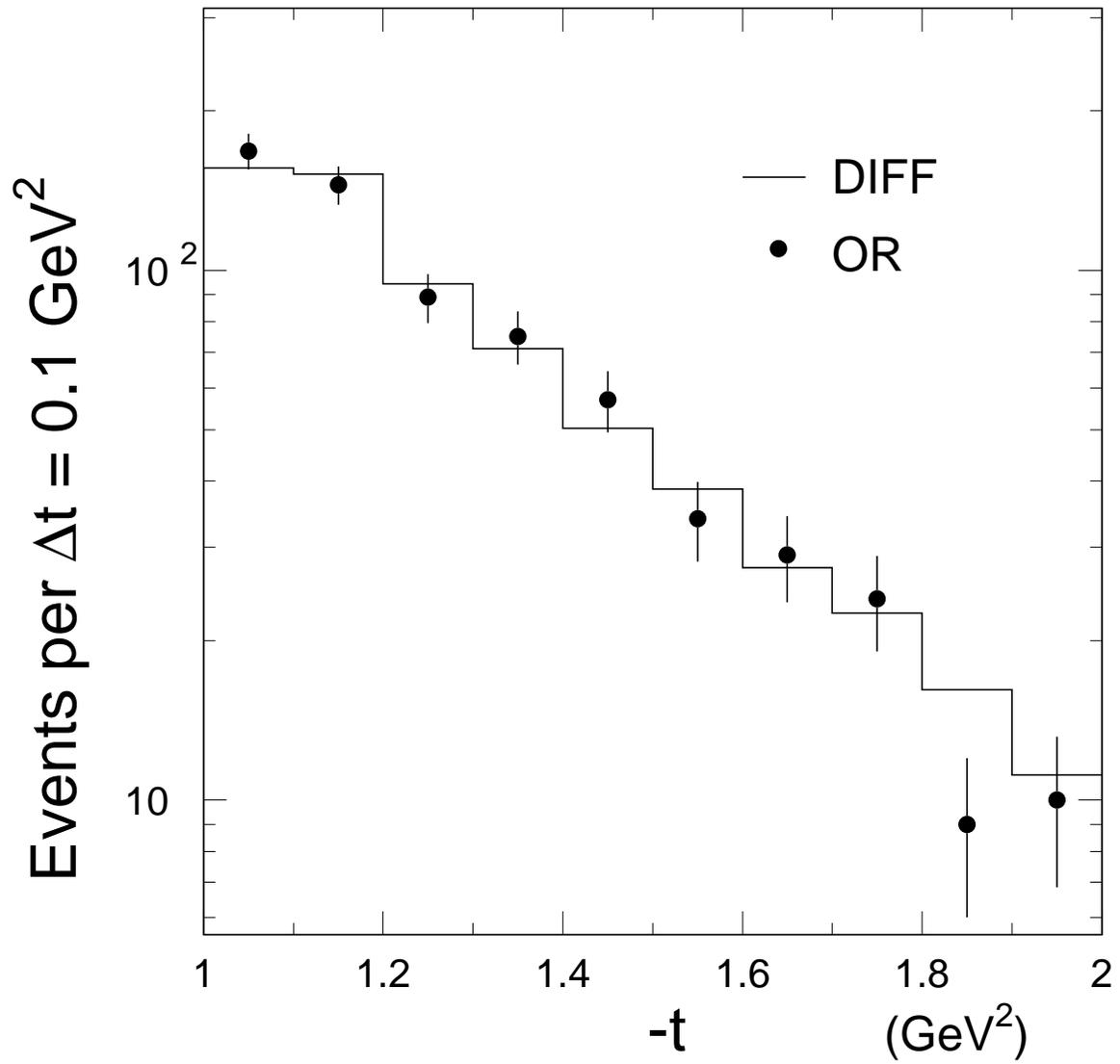,width=16cm}}
\end{center}
\caption[]{
Observed momentum-transfer distributions for 
the ``OR" data sample of React.~\ref{eq:pompom} (solid points).
The histogram normalized to the points is the single-diffractive data,
React.~\ref{eq:dif}.
}
\label{fig:tcomp}
\end{figure}

Figure~\ref{fig:tcomp} shows that the momentum transfer (\T ) 
distribution of the ``OR" events is in good agreement with that of the
full single-diffractive data sample. This is consistent with our
basic assumption that the flux factor is common to Reacts.~\ref{eq:dif}
and \ref{eq:pompom}. The lower statistics ``AND" data sample (not shown here)
is also compatible with the single--diffractive data.

\subsection{Feynman-\xp\ distribution}
\label{sect:xp}
\indent

\begin{figure}
\begin{center}
\mbox{\epsfig{file=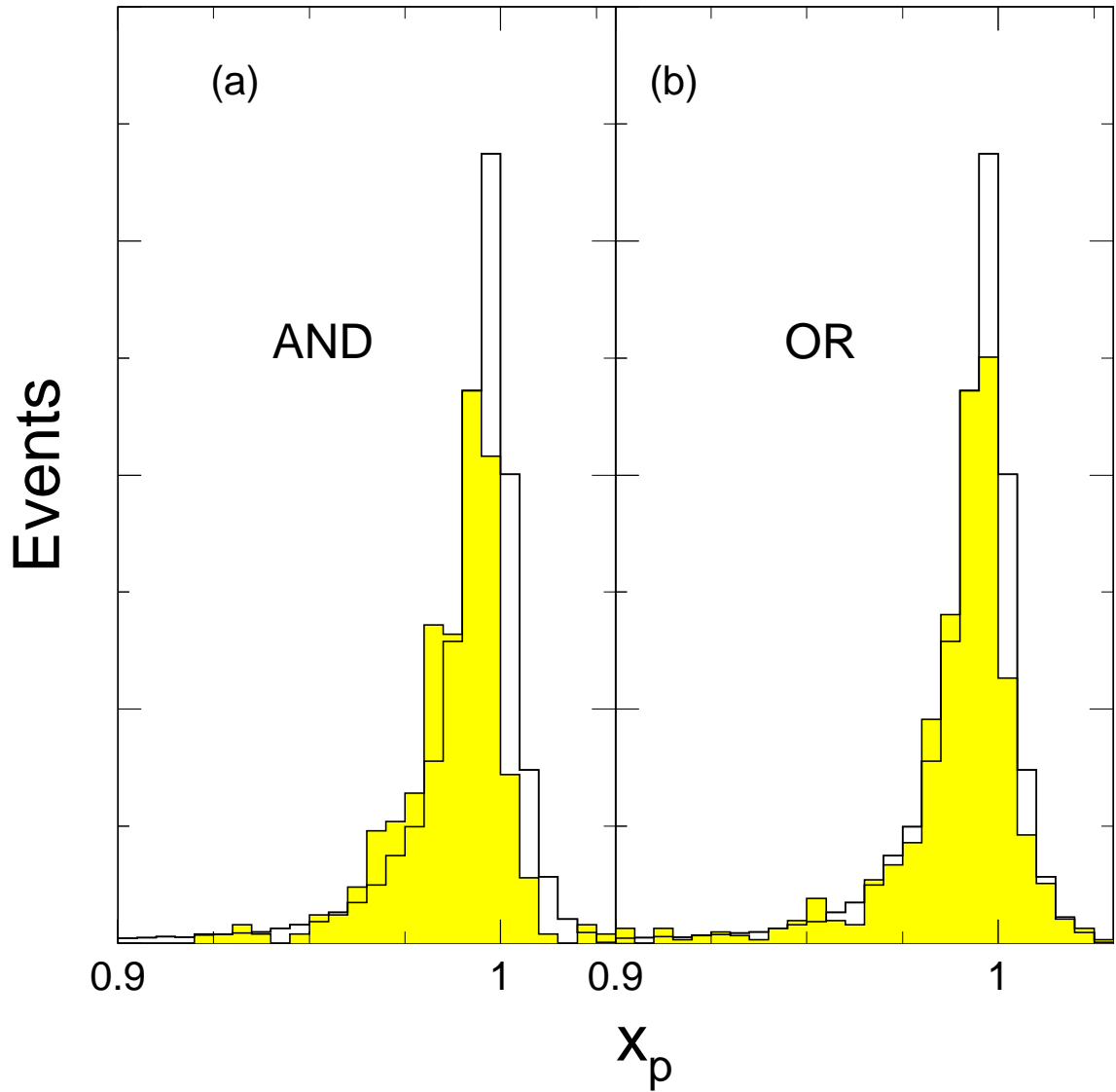,width=16cm}}
\end{center}
\caption[]{
Feynman-\xp\ distribution;
(a) Shaded histogram is \xp\ and $x_{\bar{p}}$ (2 points per event) 
for the ``AND" events of React.~\ref{eq:pompom} 
(139 events with TOF and $\Phi_X$ cuts).
The open histogram is the \xp /$x_{\bar{p}}$ distribution from inclusive
single diffraction~\protect\cite{ua8dif}, with a TOF veto
only on the trigger side.  
The shaded and open histograms are normalized to the same area for 
the bin, $0.990 < \xp < 0.995$;
(b) Same as (a), but the shaded histogram is \xp\ or $x_{\bar{p}}$
for the ``OR" data (698 events with TOF and $\Phi_X$ cuts).
The vertical scale is linear.
}
\label{fig:xcomp}
\end{figure}

The shaded distributions in Figs.~\ref{fig:xcomp}(a,b) show the distributions
of Feynman-$x_p$ and $x_{\bar{p}}$ for the final ``AND" and ``OR" data 
samples, respectively. They are essentially indistinguishable.
The open histogram superimposed on both ``AND" and ``OR" distributions 
(shaded) in Fig.~\ref{fig:xcomp} is the \xp /$x_{\bar{p}}$ distribution in the 
single--diffractive data of React.~\ref{eq:dif} in our experiment~\cite{ua8dif}.
In order that both sets of distributions have the same kinematic conditions, 
the single--diffractive data are plotted only for those events that have
no hits in the TOF counters on the trigger side, 
which cover pseudorapidity, $2.3 < \eta < 4.1$.
Each open histogram is normalized to its shaded distribution for
the bin: $0.990 < \xp < 0.995$.

We see that the single--diffractive data possess a significant event
population for $x_p > 0.995$, which is not seen in either set of data for 
React.~\ref{eq:pompom}.
As discussed in the Introduction and in Sect.~\ref{sect:mc}, 
this apparent breakdown of factorization is merely a kinematic suppression in 
React.~\ref{eq:pompom}, due to the requirement 
that $s' > 0$ for the two--\pom\ system.
In the Monte--Carlo generation of two independently--emitted \pom s
according to  Eq.~\ref{eq:pompomdist} (see Sect.~\ref{sect:mc}), 
61\% of all events, in which both $p$ and \ap\ have $|t| > 1$~GeV$^2$,
have $s' < 0$ and are discarded. 
The rejected events mostly have small--$\xi$; for example, 
when either \pom\ has
$\xi < 0.0005$, 100\% of the events are rejected, whereas when 
$\xi \approx 0.03$, only 26\% of the events are rejected. 
This qualitatively accounts for the
difference between DPE and single diffractive data near \xp\ = 1
in Fig.~\ref{fig:xcomp}.
For the ``OR" topology, when only one $|t| > 1$~GeV$^2$, 
the rejection at small--$\xi$ is about the same (95\%), whereas there is
more rejection at $\xi = 0.03$ (51\%); this accounts for the small 
difference in
shape between ``AND" and ``OR" data in Fig.~\ref{fig:xcomp}. The detailed 
shapes of these distributions depend on \sigpompom , which we have not yet
determined.

\section{Monte--Carlo event generation}
\label{sect:mc}
\indent

A complete Monte--Carlo simulation~\cite{thesis} of React.~\ref{eq:pompom} 
was performed to determine the spectrometer and calorimeter acceptances
as well as the efficiencies of the various cuts.
Events were generated such that the \pom s are emitted independently from 
proton and antiproton, respectively, according to Eq.~\ref{eq:pompomdist}, 
using the \pom\ flux factor~\cite{es2,ua8dif} in Eq.~\ref{eq:dsigdif}.
\sigpompom\ is assumed to be independent of $s'$, 
although in Sect.~\ref{sect:sigma} we will look for departures from this
assumption.

Points were chosen randomly in the 6-dimensional 
space\footnote{
Since the 3 observables of a final-state proton or antiproton are uniquely
related to those of its associated \pom , we use the \pom\ variables,
$\xi$, $t$ and $\phi$.}, 
($t_1, \, \xi_1, \, \phi_1, \, t_2, \, \xi_2, \, \phi_2$), according to
the product of two flux factors. Each such point defines the properties
of the \pom --\pom\ system, its energy and its momentum vector.
We have observed that, even though the two \pom s are assumed to be
independently emitted,
not all points in the 6--dimensional space are kinematically allowed
because the associated \pom -\pom\ invariant mass may be unphysical 
(i.e., $s' < 0$).
Thus, events are retained only if they are in regions of the 6-dimensional
space for which $s' > 0$. 
In our $|t|$-domain, 1--2~GeV$^2$, such events are 39\% of the total generated. 
We note that, even though
the \pom s are generated isotropically and independently 
in azimuthal angle, $\phi$, 
correlations occur due to this kinematic suppression.

The number of particles of the central system, $X$, in React.~\ref{eq:pompom} 
is generated according to a Poisson distribution with its mean charged particle
multiplicity depending 
on $M_X$, as measured in a study of low-mass diffractive systems~\cite{bc300}, 
$\bar{n} = 0.6 M_X$ ($M_{X} = \rsp$ in GeV);
the mean number of neutral particles is
assumed to be one-half the number of charged particles. 
The tracks are generated isotropically 
in the $M_X$ center-of-mass (see Sect.~\ref{sect:ua2cal}).
As described in Sect.~\ref{sect:central}, where the central system is seen
to have longitudinal structure for $M_X > 5$~GeV, the Monte--Carlo generator 
is tuned to agree with the data.

After phase-space generation of the complete events,
their data were passed through detector simulation software for both
the UA8 spectrometers and the UA2 detectors~\cite{ua2sim}, and then through
the same offline analysis software and cuts used for the real data.

As already discussed in Sect.~\ref{sect:evsel},
Table~\ref{tab:mc} shows the good 
agreement between the real event losses with the 
momentum cuts described in Sect.~\ref{sect:evsel}
and those on the Monte--Carlo events calculated here. 
Since the Monte--Carlo
event sample of React.~\ref{eq:pompom} suffers a 46\% loss
when the TOF veto is imposed, the net efficiency for event retention
due to TOF veto and event selection cuts is 26\%.

\begin{figure}
\begin{center}
\mbox{\epsfig{file=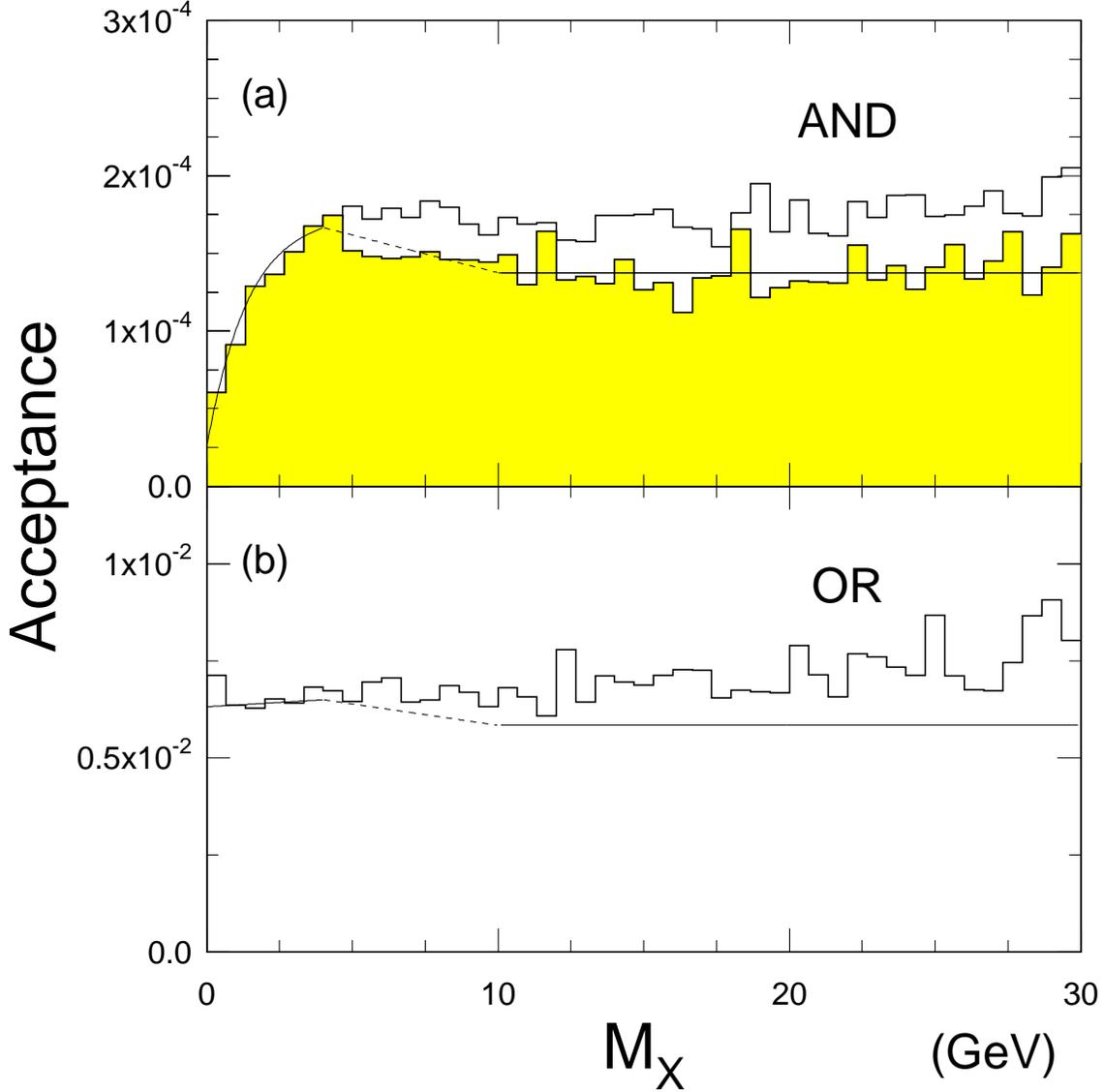,width=16cm}}
\end{center}
\caption[]{
Geometrical acceptance and track reconstruction efficiency
vs. $M_X$. As described in the text, retention efficiencies of
TOF veto cut and event selection are included;
(a) ``AND" triggered events with $p$ and \ap\ both having
$1.0 < -t < 2.0$~GeV$^2$.
As discussed in the text, the open histogram assumes isotropic decay
and the shaded histogram assumes longitudinal decay, for $M_X > 4$~GeV.
The combination of solid and dashed curves is the acceptance function
used in cross section calculations; 
(b) ``OR" triggered events with $1.0 < -t < 2.0$~GeV$^2$ for the observed
final-state particle assuming isotropic decay. 
The solid line for $M_X > 10$~GeV is the assumed acceptance
for longitudinal decay which, as for the ``AND" data,
is $\approx 25\%$ smaller than the acceptance for events with 
isotropic decay. 
}
\label{fig:mcacc}
\end{figure}

The combined geometric and detection efficiency of proton and antiproton 
is about $6 \cdot 10^{-4}$ at an average $|t|$ of 1.2~GeV$^2$. 
Figures~\ref{fig:mcacc}(a,b) show the $M_X$--dependence of the overall 
geometric and detection efficiencies averaged over all other variables, 
for ``AND" and ``OR" data, respectively, when the observed particles are 
in the range, $1.0 < -t < 2.0$~GeV$^2$. The 26\% central
system detection efficiency is also included in these efficiencies.
The fall--off in acceptance for the ``AND" data at low mass results from the 
fact that, kinematically, low--mass events tend to have back--to--back
$p$ and \ap , which do not satisfy the ``AND" trigger topology seen in
Fig.~\ref{fig:dpevector}.  
The ``OR" trigger topology does not have such a bias against low--mass events.

The longitudinal structure seen in Sect.~\ref{sect:central} for central
system masses larger than about 5 GeV must impact the 
acceptance of the central system.
Thus, we show the ``AND" acceptances in Fig.~\ref{fig:mcacc}(a),
first assuming isotropic decay of the central system and then, 
for $M_X > 4$~GeV, 
with a longitudinal decay distribution which matches that which is observed. 
At large mass, the acceptance for longitudinal decay is 25\% smaller than 
the acceptance for isotropic decay.
Since, with our statistics, it is not possible to properly study the
transition from isotropic to longitudinal decay, we take the effect into
account in the following way.
For $M_X < 4$~GeV, we use the calculated acceptance for isotropic decay, 
shown as the solid curve.
For $M_X > 10$~GeV, we use the fitted horizontal solid line in the figure. 
In the intermediate range, between 4 and 10 GeV, the dotted interpolation line 
is used.

For the ``OR" acceptance at low mass ($M_X < 4$~GeV), we use the calculated 
acceptance for isotropic decay in Fig.~\ref{fig:mcacc}(b),
just as we did for the ``AND" data. 
For $M_X > 10$~GeV,  we use the solid horizontal line which is 
25\% below the calculated efficiency for isotropic decay. Again, the
dashed interpolation line is used in the intermediate region. 

\section{Calorimeter measurement of central system}
\label{sect:ua2cal}
\indent

We use the UA2 calorimeter information to study the invariant mass and
other properties of the central system, $X$, in React.~\ref{eq:pompom}.
The UA2 detector simulation software~\cite{ua2sim} was used to perform a 
complete Monte--Carlo study of the calorimeter response. As noted above,
the UA2 simulation software is remarkably good in describing
the low-energy deposits encountered in our data.

\subsection{Invariant mass of central system}
\label{sect:mass}
\indent 

Since we do not directly observe the individual particles of this system,
but rather the energies deposited in the calorimeter cells,
we assume that the non-zero energy in each ``struck" cell of the calorimeter
is caused by a massless particle, and then calculate,
\begin{equation}
{M_{observed}}^2 = 
(\Sigma E_i)^2 - |\Sigma \vec{P}_i|^2, 
\label{eq:mass}
\end{equation}
summing over all cells.

\begin{figure}
\begin{center}
\mbox{\epsfig{file=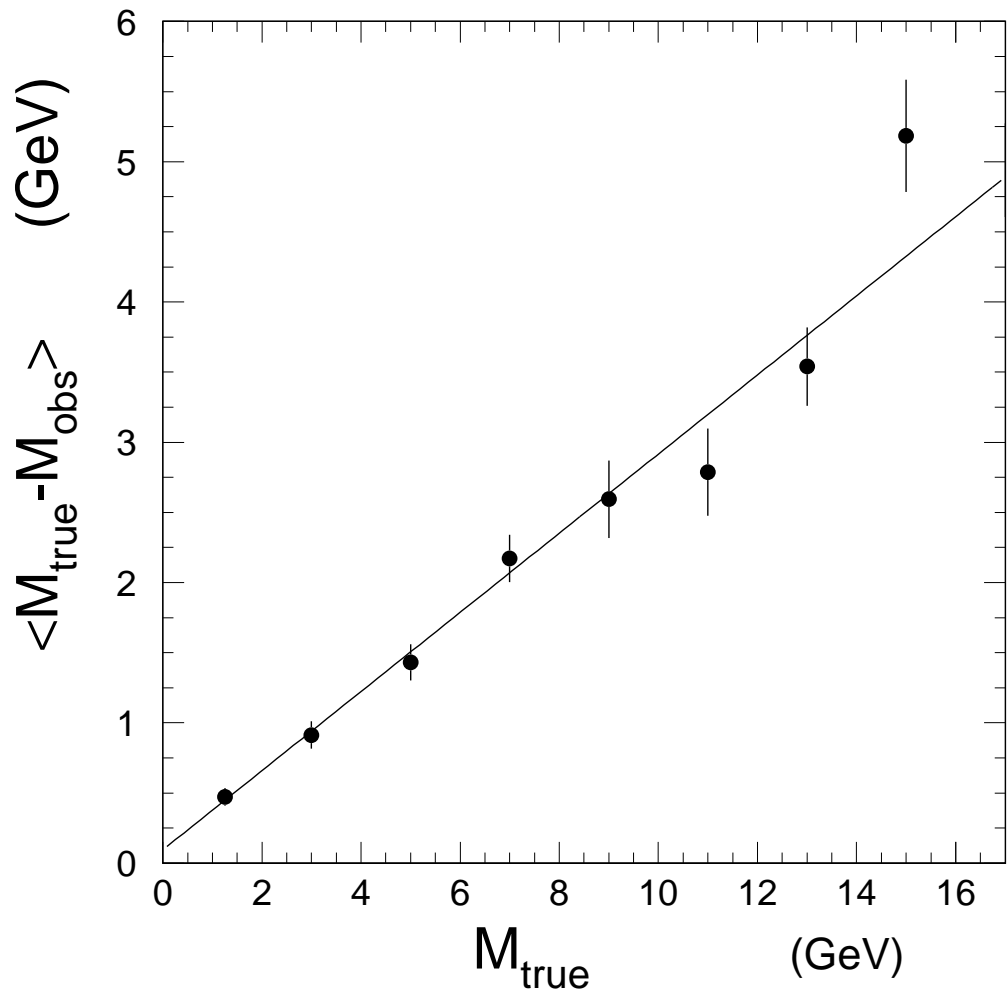,width=14cm}}
\end{center}
\caption[]{
Monte--Carlo study of invariant mass calculation using the calorimeter.
The observed (downward) shift in mass (true - observed) vs. the true
mass. The fitted line corresponds to Eq.~\ref{eq:correction} in the text.
}
\label{fig:masscor}
\end{figure}

Figure~\ref{fig:masscor} is the result of a Monte--Carlo study~\cite{thesis}
which shows that this procedure underestimates the true
mass by an amount that increases with mass. 
This effect results from incomplete detection of energy;
for example, the finite cell size leads to 
overlapping energy deposits from neighboring particles,
or some energy can be lost before the particles enter the calorimeter.
The difference between the
true MC mass, $M_{true}$, and the calculated or observed mass, 
$M_{observed}$, is plotted vs. $M_{true}$. 
The dependence is well fit by the equation (with $M$ in units of GeV): 
$M_{true} - M_{observed} = (1 + 4 M_{true}) / 14$, which can be rewritten as:
\begin{equation}
M_{true} \, \, = 1.4 \, M_{observed} \, \, + \, \, 0.1.
\label{eq:correction}
\end{equation}
We define the corrected mass, $M_{X}^2 = s'$, to be the true mass given by 
this equation and only refer to these corrected values in the remainder of 
this paper.

\begin{figure}
\begin{center}
\mbox{\epsfig{file=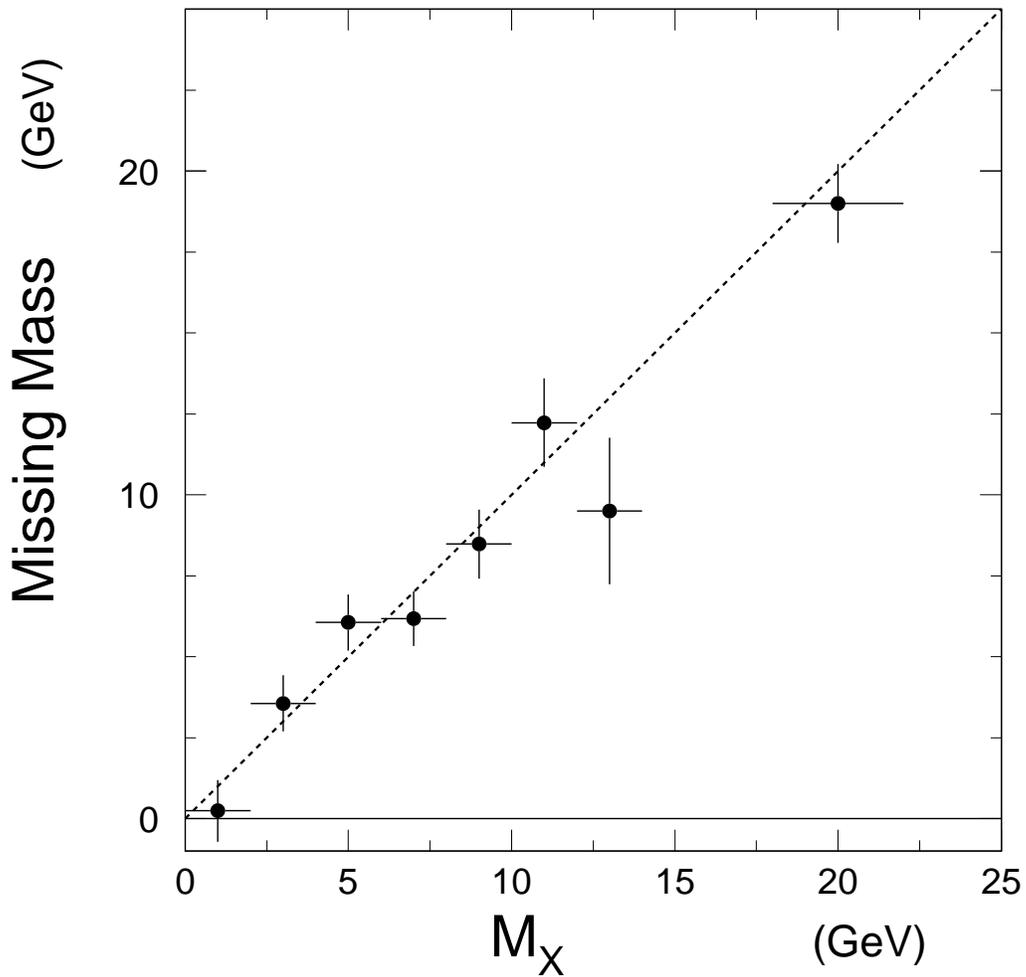,width=14cm}}
\end{center}
\caption[]{
The mean ``missing mass" calculated from the observed $p$ and \ap ,
vs the corrected invariant mass calculated from the calorimeter 
information. See discussion in text.
The vertical error bars on each point are the errors-in-the-mean
for the missing mass calculation, while the horizontal bars show
the event binning.
As seen in Fig.~\ref{fig:dndmx}, there are zero events in the
$M_X$ range 14--18 GeV.
}
\label{fig:pmm}
\end{figure}

The validity of the calorimeter invariant mass calculation may be 
conveniently tested
by comparing it with the {\it missing} mass calculated using the measured
$p$ and $\bar{p}$ 4-vectors for an event.
Although the experimental uncertainty in a ``missing mass" 
calculation is much larger than for the calorimeter invariant mass,
they should agree on average.
Fig.~\ref{fig:pmm} shows the average missing mass calculated for the events 
in each of the calorimeter invariant mass bins shown in the figure.
The observed clustering of the 
points around the diagonal and the absence of any systematic shifts
constitutes proof that the calorimeter mass evaluation is reliable.

The $M_X$ distributions of the system $X$ in React.~\ref{eq:pompom} are shown 
for the final selected ``AND" and ``OR" event samples in 
Figs.~\ref{fig:dndmx}(a,b), after requiring that the momentum transfer
of all detected protons and antiprotons be in the range, 1--2 GeV$^2$.
From the relatively flat 
acceptance curves in Figs.~\ref{fig:mcacc}(a,b), we see that 
the observed shapes of the distributions are reasonably good
representations of the true distributions (except for the lowest mass bin in 
the ``AND" data) .
In Sect.~\ref{sect:ppsigma}, we show that part of the low--mass 
enhancements are attributable to an explicit $s'$ dependence in \sigpompom ,
corresponding to an enhanced \pom --\pom\ interaction in the few-GeV
mass region.
However, with an estimated 1.8 GeV mass resolution obtainable from
the calorimeter, we are unable to observe details of any possible 
$s$-channel resonant structure in this spectrum.  

\subsection{Other properties of the central system}
\label{sect:central}
\indent

\begin{figure}
\begin{center}
\mbox{\epsfig{file=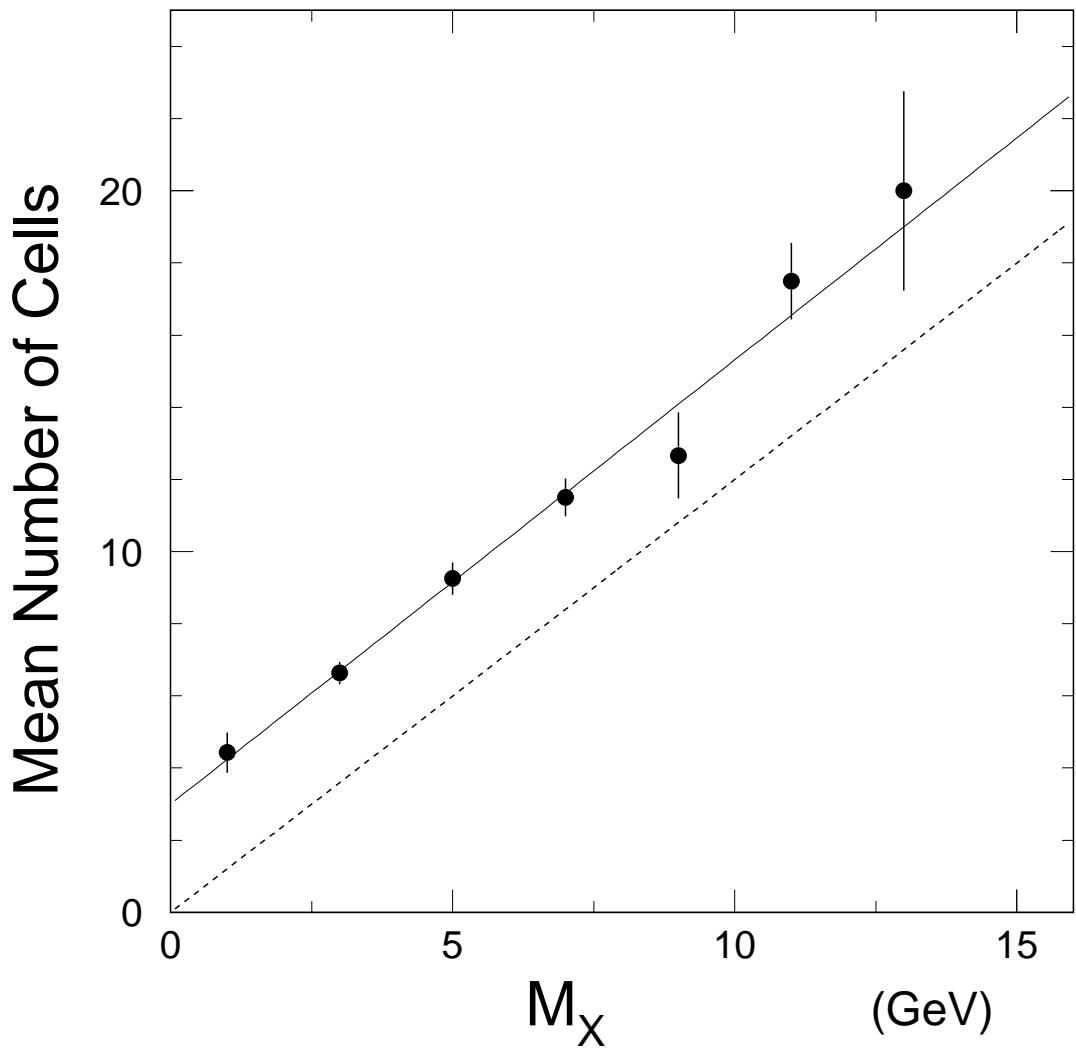,width=15cm}}
\end{center}
\caption[]{
Mean number of struck calorimeter cells with at least 200~MeV energy 
(total energy in electromagnetic and hadronic sections) vs. 
corrected calorimeter invariant mass. 
Dashed line is the naive expectation, using $<N> = 1.2 M_X$, as discussed
in the text.
}
\label{fig:ncell}
\end{figure}

In addition to the invariant mass distribution of the system, $X$, other
properties of the system can be studied. One is the particle multiplicity
of the central system.
Figure~\ref{fig:ncell}
shows the number of calorimeter cells struck as a function of the
corrected calorimeter mass, $M_X$.
The solid line is a fit to the data;
the dashed line is based on the naive multiplicity expectation 
assuming~\cite{bc300} $<N> = 0.6 M_X$ ($M_X$ in GeV)
for the number of charged particles 
($\pi^+$ and $\pi^-$). 
The number of $\pi^0$ is assumed to be Poisson--distributed with a mean of
0.3$M_X$; each of these is assumed to appear as two $\gamma$. 
The resulting dashed line is the function $N = 1.2 M_X$ and
clearly captures the gross features of the data.
The observed numbers of struck cells lie somewhat above the line, as expected 
geometrically
from the cluster widths in the calorimeter and the finite cell sizes.
A complete Monte--Carlo simulation~\cite{thesis} 
accounts for the small observed differences.
We can conclude that the number of observed struck cells 
increases with mass roughly as expected, and the total observed multiplicity
displays no anomalous features.

\begin{figure}
\begin{center}
\mbox{\epsfig{file=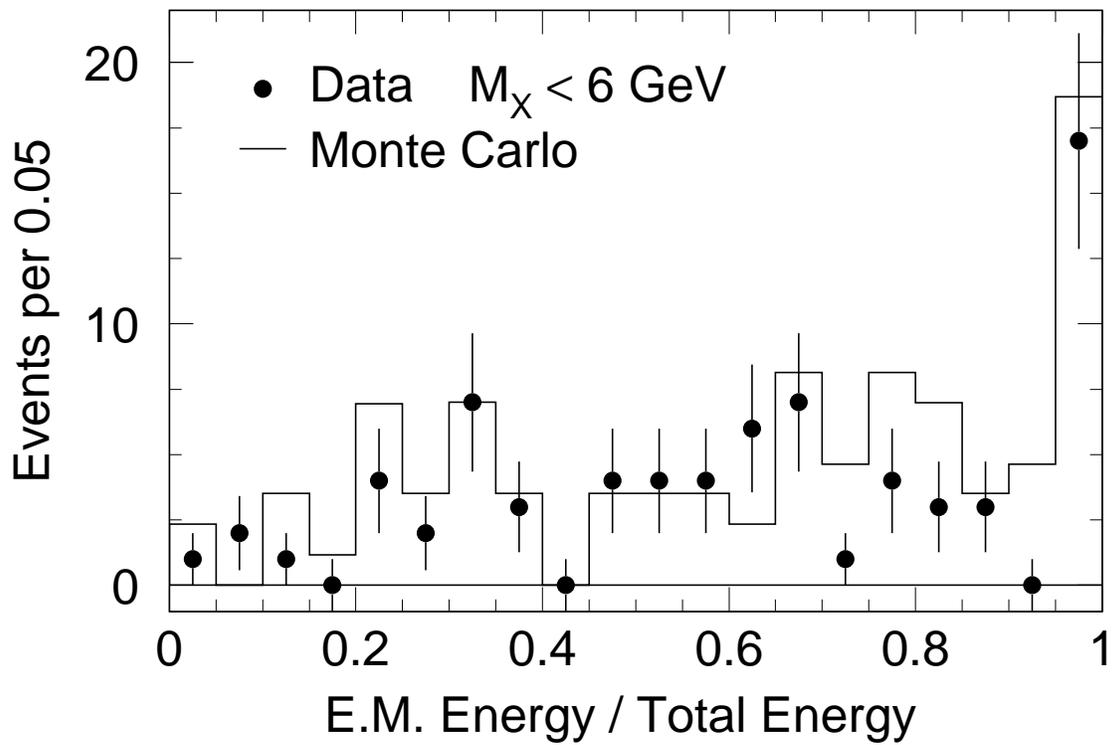,width=16cm}}
\end{center}
\caption[]{
Number of ``AND" events with $M_X < 6$~GeV vs. ratio of electromagnetic energy 
to total energy detected in calorimeter.
The peak corresponding to (e.m. energy = total energy) in both data and 
Monte Carlo is due to low--energy charged tracks losing all their energy
in the e.m. sections of the UA2 calorimeter.
}
\label{fig:pem}
\end{figure}

Because of the separated electromagnetic and hadronic section of the
UA2 calorimeter,
it has also been possible to study the fraction of electromagnetic energy
possessed by the central system. Fig.~\ref{fig:pem} shows the distribution
in this fraction for the ``AND" 
events in the low-mass enhancement, $M_X < 6$~GeV,
compared with the Monte--Carlo generated distribution 
assuming, on the average,
equal numbers of $\pi^{+}$, $\pi^{-}$ and $\pi^{0}$ in the track generation.
Again we see no anomalous features in this variable.
The enhancement visible in both data and Monte--Carlo when the ratio, 
(e.m. energy)/(total energy), equals unity corresponds to low mass systems 
where the slow pions deposit all their energy 
in the electromagnetic calorimeter cells.

\begin{figure}
\begin{center}
\mbox{\epsfig{file=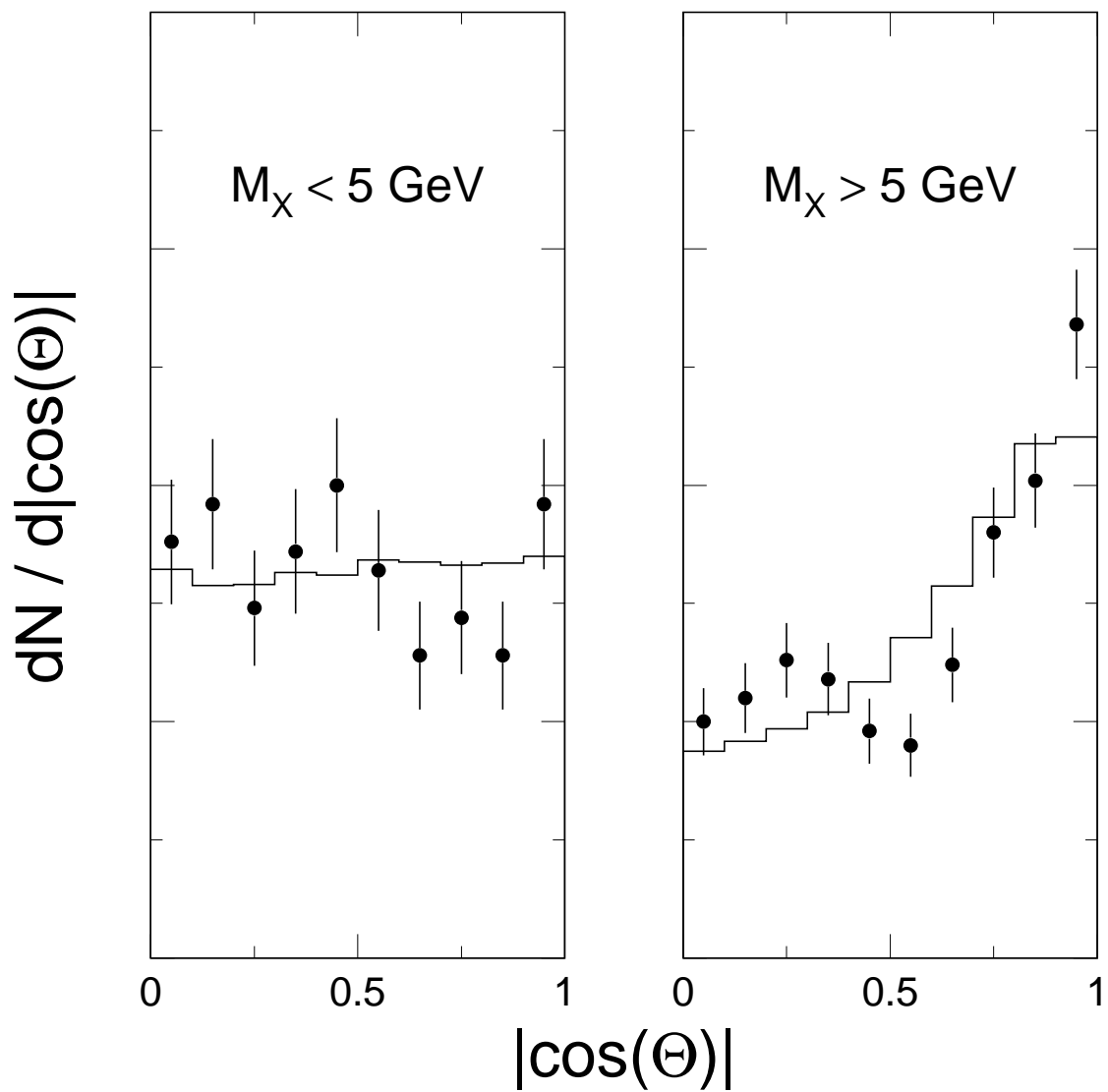,width=16cm}}
\end{center}
\caption[]{
Central system decay distributions.
$dN/dcos(\theta )$ for all ``struck'' cells, averaged over
the event sample: 
(a) for $M_X < 5$~GeV; 
(b) for $M_X > 5$~GeV.
Histograms are the Monte--Carlo distributions described in the text.
Vertical scale is arbitrary and linear.
}
\label{fig:costh}
\end{figure}

We have examined the angular distributions of calorimeter cell
energies in the center-of-mass of the $X$ system, $dN/dcos\theta$,
with respect to the \pom -\pom\
direction of motion. Figs.~\ref{fig:costh}(a,b) show these distributions for
$M_X < 5$ GeV and $M_X > 5$ GeV. 
At the higher masses we see a similar type of forward--backward peaking 
as is seen in all hadronic interactions as a result of the presence of
spectator partons. 
In the present case, this would imply that there are
spectator partons in the \pom .
We have already reported~\cite{ua8dif,mike} 
similar effects in \pom -proton interactions in the
single-diffractive, React.~\ref{eq:dif}.

The Monte--Carlo histogram in Fig.~\ref{fig:costh}(a) shows 
isotropically-decaying events. 
In Fig.~\ref{fig:costh}(b), the histogram shows a Monte Carlo event sample
which has been selected in such a way that it has the same forward--backward
peaking as the experimental distribution.
For each isotropically--decaying Monte--Carlo event, the mean value of 
$cos^2 \theta$ is evaluated 
averaging over all outgoing tracks in the central system. 
We have found \cite{thesis} that, if Monte--Carlo
events are selected for which this quantity is larger than 0.375,
the selected events follow the experimental $cos \theta$ distribution.

\section{Cross sections}
\label{sect:sigma}
\indent 

\begin{figure}
\begin{center}
\mbox{\epsfig{file=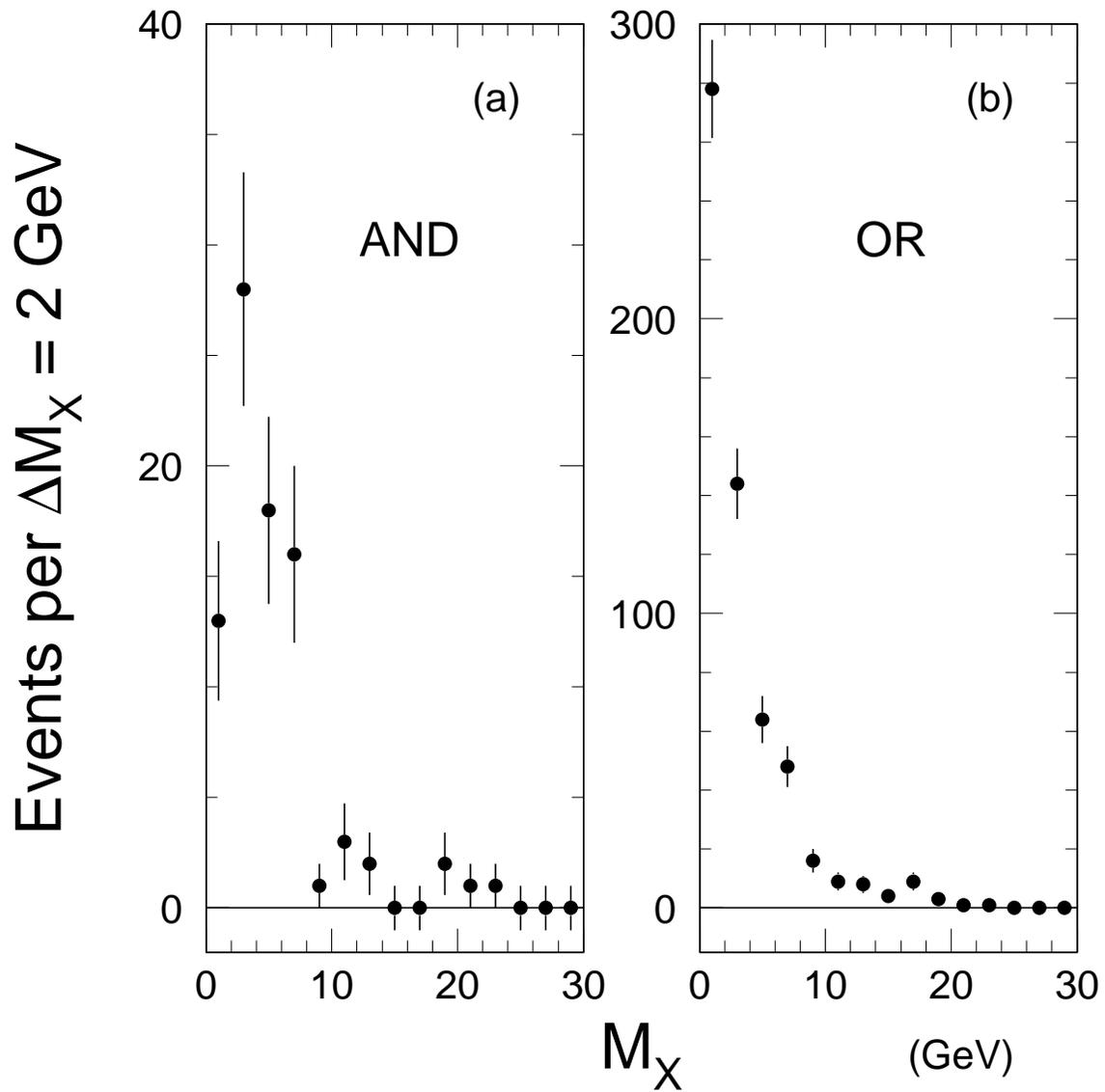,width=16cm}}
\end{center}
\caption[]{
Final event sample; number of observed events vs. corrected calorimeter mass,
$M_X$, with $1.0 < -t < 2.0$~GeV$^2$;
(a) ``AND" triggered data (85 events);
(b) ``OR" triggered data (586).
}
\label{fig:dndmx}
\end{figure}

The observed mass distributions, $dN/dM_X$, 
shown in Figs.~\ref{fig:dndmx}(a,b)
for the ``AND" and ``OR" data, respectively, are converted into absolute
cross section distributions, $d\sigma_{DPE}/dM_X$, in the following way.
Bin--by--bin, the numbers of events in Figs.~\ref{fig:dndmx}
are divided by the Monte--Carlo acceptance curves in 
Figs.~\ref{fig:mcacc}(a,b). 
Then, all are divided by a global efficiency, $\epsilon_0$,
for event retention when halo and pileup cuts~\cite{ua8spect, thesis} are made, 
and by the appropriate integrated luminosity for each trigger sample: 
\begin{equation}
\Delta \sigma_{DPE} \, \, = \, \, 
{{\Delta N}\over{\int {\cal L} dt \x \epsilon_0 \x A}}.
\label{eq:crsec}
\end{equation}

The ``AND" (``OR") data samples have an efficiency, $\epsilon_0$, 
of 0.54 (0.76) and an integrated luminosity, $\int {\cal L} dt$,
of  2894~nb$^{-1}$ (5.4~nb$^{-1}$).
The ``OR" luminosity is ``effective", due to prescaling of the ``OR" trigger.
In both cases, the cross section is given only for momentum transfer
of the observed trigger particle(s) in the range:  $1.0 < |t| <2.0$~GeV$^2$.
In the ``OR" case, the unseen $p$
or \ap\ has its ``natural" \pt\ distribution and therefore peaks at small
values. Thus, the observed cross section is much larger for the ``OR" data. 
The resulting cross sections for the two triggered data samples
of Reaction~\ref{eq:pompom},
$d\sigma/dM_X$, are the points shown in Figs.~\ref{fig:dsigdmx}(a,b).

\subsection{\boldmath \pom --\pom\ total cross section}
\label{sect:ppsigma}
\indent

\begin{figure}
\begin{center}
\mbox{\epsfig{file=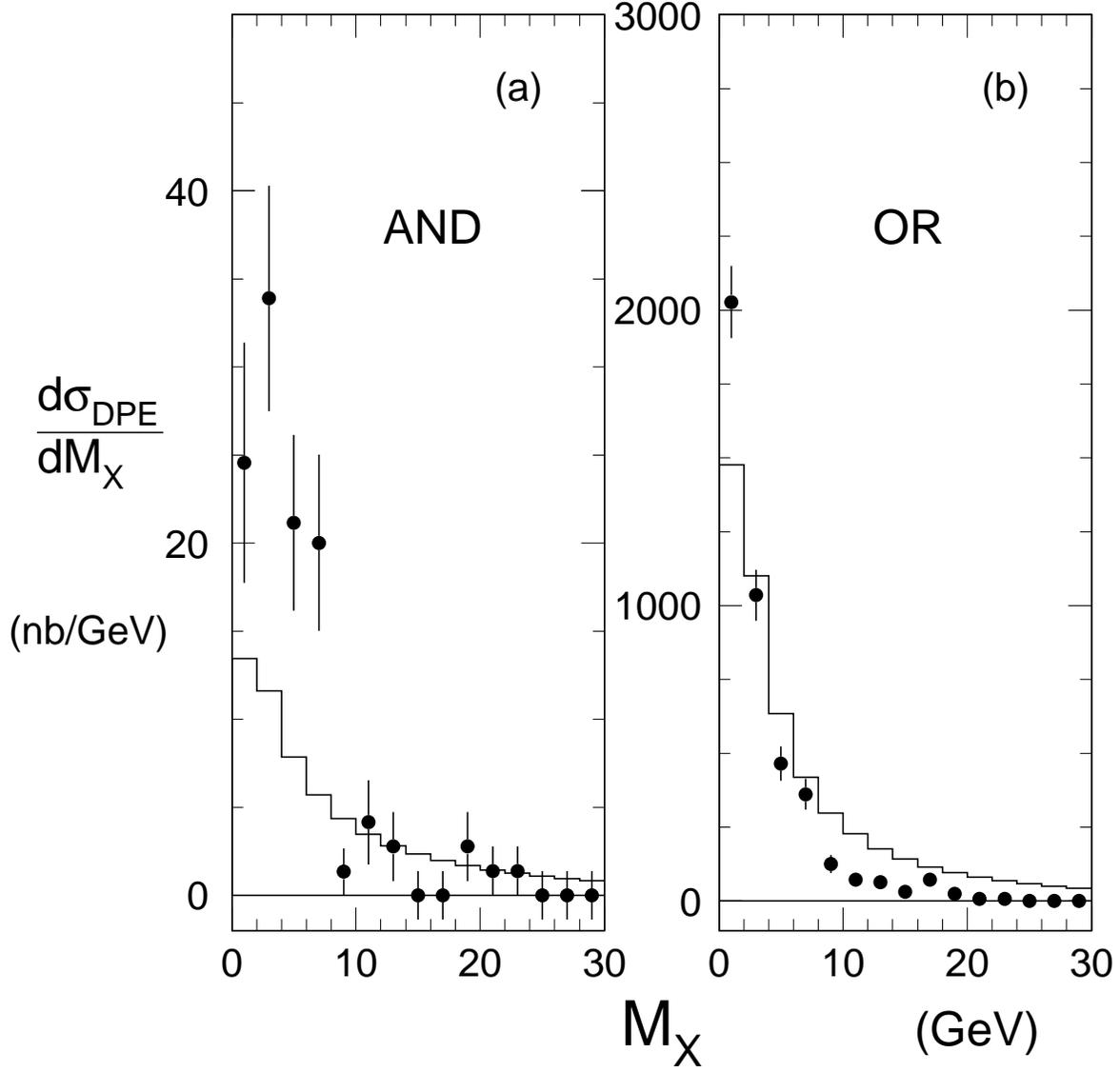,width=16cm}}
\end{center}
\caption[]{$d\sigma_{DPE} / dM_X$, the
corrected differential cross section for React.~\ref{eq:pompom} 
(proportional to the ratios of Figs.~\ref{fig:dndmx} and Figs.~\ref{fig:mcacc}),
only for momentum transfer(s), \T , of the observed trigger 
particle(s), $p$ and/or \ap , in the range, $1.0 < -t < 2.0$~GeV$^2$;
(a) ``AND" triggered data;
(b) ``OR" triggered data.
The histograms are Monte--Carlo predictions assuming $M_X$--independent,
$\sigpompom = 1$~mb. 
As discussed in the text, the absolute values shown assume the 
(somewhat arbitrary) value, $K = 0.74$~GeV$^{-2}$. 
}
\label{fig:dsigdmx}
\end{figure}

We now extract the \pom --\pom\ total cross section,  \sigpompom ,  
from the data, so that we can look for deviations from our
earlier assumption that it is independent of $M_X$.
The histograms in Figs.~\ref{fig:dsigdmx}(a,b) are Monte--Carlo 
predictions for the $d\sigma_{DPE}/dM_X$ points in the figure,
made using Eq.~\ref{eq:pompomdist} and \flux\ in Eq.~\ref{eq:dsigdif}.
We assume a constant \pom --\pom\ total cross section of \sigpompom\ = 1~mb
and the (arbitrary) value, $K = 0.74$~GeV$^{-2}$, as discussed in
Sect.~\ref{sect:intro}.

Since the UA8 spectrometer acceptances~\cite{ua8dif} do not vary
significantly over the range of $\xi$ studied here,
the ratios of the points to the histogram values in Figs.~\ref{fig:dsigdmx}
give values of the \pom -\pom\ total cross section, \sigpompom , 
vs. $M_X$. 
These ratios are shown in Fig.~\ref{fig:sigpp},
for both the ``AND" and ``OR" data.
We note that, despite the large difference between the measured cross 
sections, $d\sigma_{DPE} / dM_X$, in Figs.~\ref{fig:dsigdmx}(a,b), 
both data sets yield the same general properties for 
\sigpompom : enhancements for $M_X < 8$~GeV and relatively $M_X$ independent
shapes at larger $M_X$. 
Although the two values of \sigpompom\ 
in the first bin ($M_X < 2$~GeV) are consistent with being equal, in the 
next three bins ($2 < M_X < 8$~GeV) the ``AND" cross sections are about
three times larger than the ``OR" cross sections. 
However, only the statistical errors on \sigpompom\ 
are shown in Fig.~\ref{fig:sigpp}. We discuss their systematic 
uncertainties in the following section.

\begin{figure}
\begin{center}
\mbox{\epsfig{file=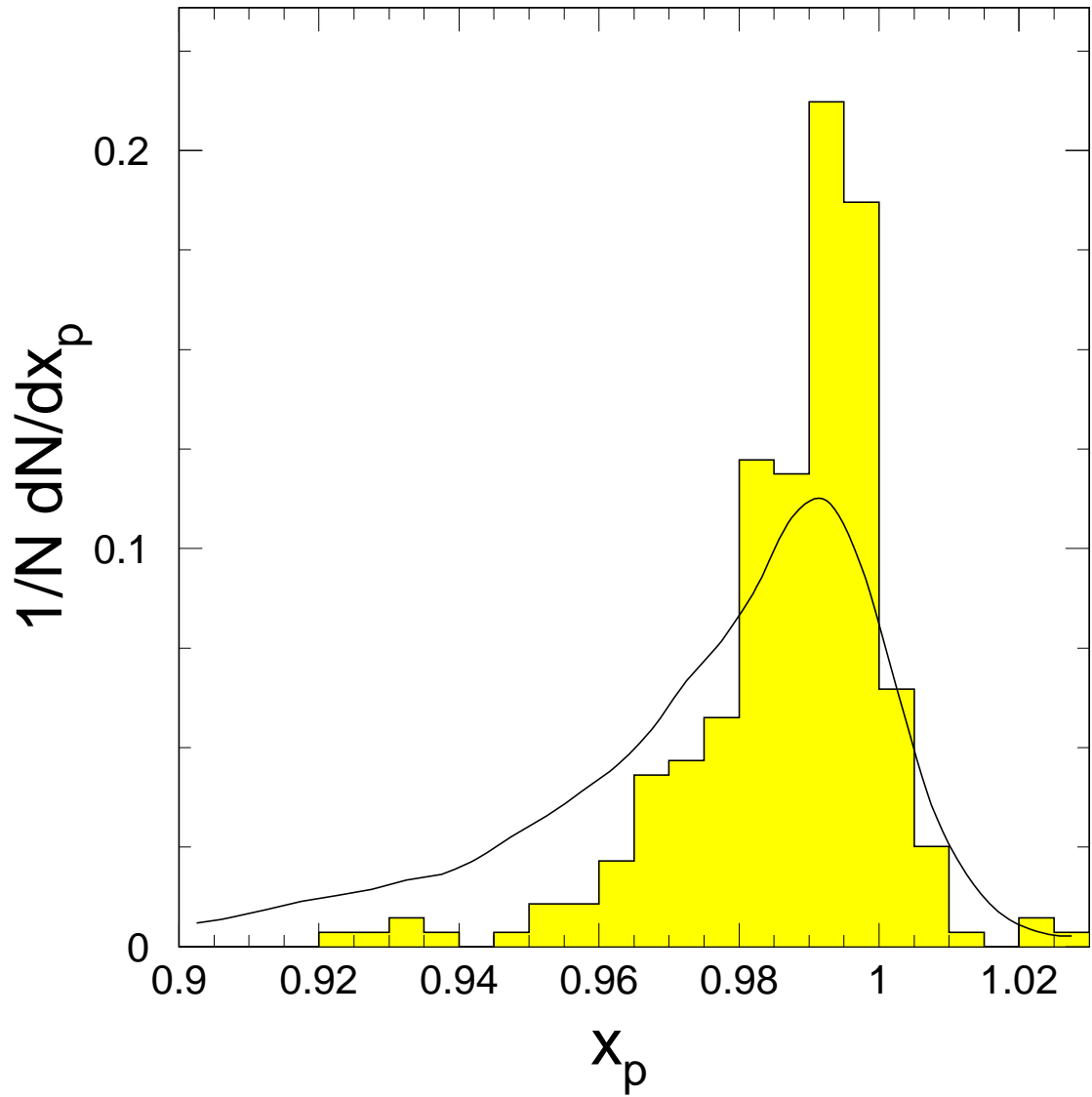,width=16cm}}
\end{center}
\caption[]{
\xp\ ($x_{\bar{p}}$) 
distribution for the ``AND" data as in Fig.~\ref{fig:xcomp}(a).
The solid curve normalized to the data 
is the Monte--Carlo prediction assuming no explicit 
$s'$--dependence in \sigpompom . 
}
\label{fig:xpmc}
\end{figure}

The small--$M_X$ enhancement in Fig.~\ref{fig:sigpp} 
also reflects itself in the 
observed \xp\ ($x_{\bar{p}}$) 
distributions seen above in Figs.~\ref{fig:xcomp}. 
Such a correlation must exist because of the kinematic relation,
$ {M_X}^2 = s' = \xi_1 \xi_2 s$; small $M_X$ correlates with small $\xi$. 
Fig.~\ref{fig:xpmc} repeats the \xp\ ($x_{\bar{p}}$) distribution 
for the ``AND" data in Fig.~\ref{fig:xcomp}(a). 
The solid curve normalized to its area is the
Monte--Carlo prediction which assumes an $s'$--independent \sigpompom .
The pronounced excess of events near \xp\ = 1.0 in the experimental 
distribution, compared with the Monte--Carlo distribution, 
is another manifestation of the low--mass enhancement in \sigpompom .

The low--mass enhancements seen in both distributions in 
Figs.~\ref{fig:sigpp} are most likely too large\cite{pvl} to be due to
a breakdown of factorization at small mass, especially for the ``AND" data.
Thus, the rise may indicate that glueball production
is a significant component of the low-mass \pom -\pom\ interaction,
although not necessarily an $s$-channel effect. That is, the observed 
invariant mass could be that of a glueball $\it plus$ other particles,
which would not lead to resonance structure in the mass distribution.
In any case, with a mass resolution of $\approx 1.8$~GeV, 
no $s$--channel structure could be seen.

\subsection{Systematic uncertainties}
\label{sect:systematic}
\indent

\begin{figure}
\begin{center}
\mbox{\epsfig{file=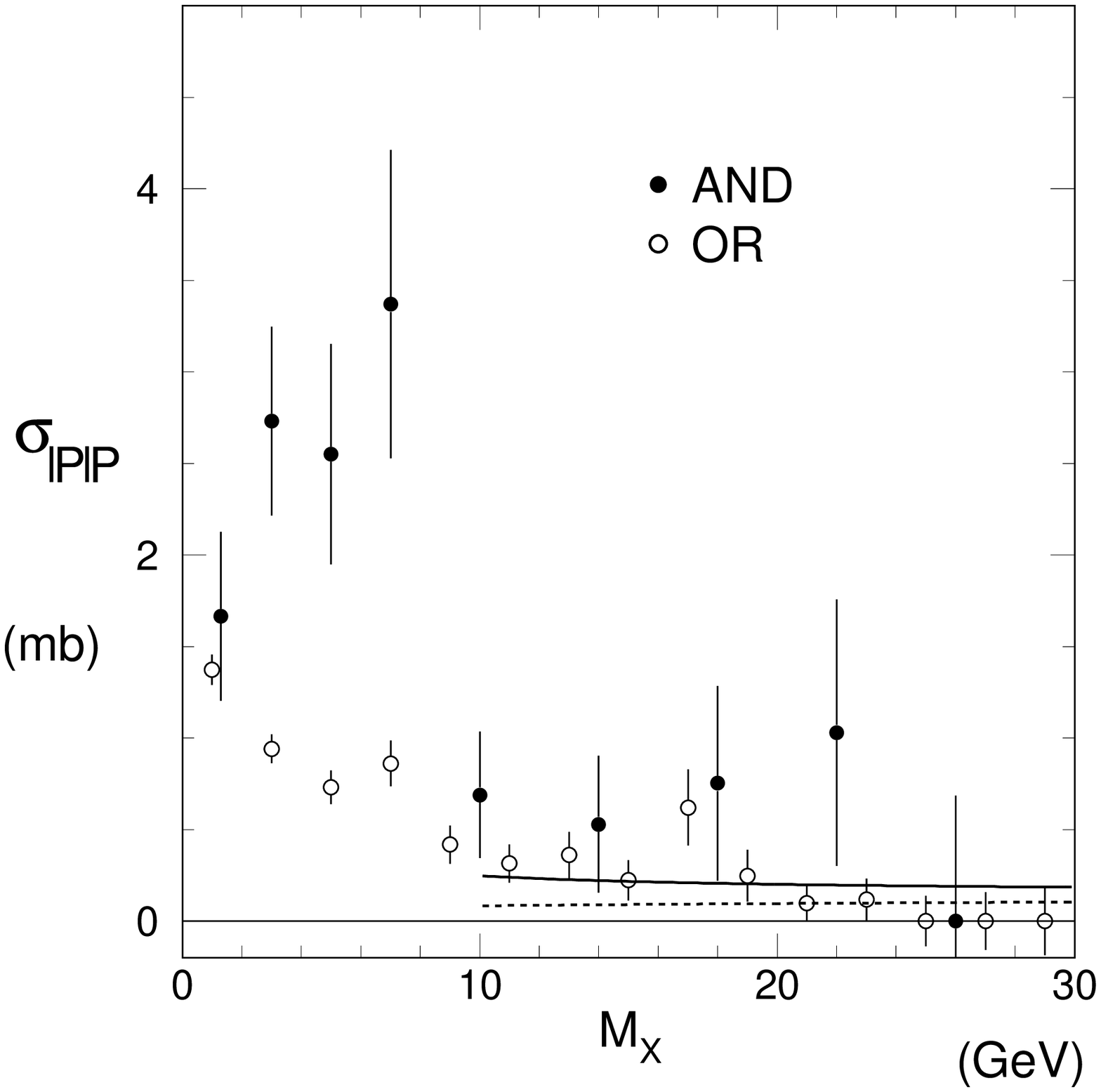,width=16cm}}
\end{center}
\caption[]{
Mass dependence of the \pom --\pom\ total cross section, \sigpompom ,
derived from the ``AND"  and ``OR" triggered data, respectively.
The (arbitrary) cross section scale assumes $K = 0.74$~GeV$^{-2}$,
as explained in the text.
Dashed curve is the factorization prediction (which is independent
of the assumed value of $K$) for the \pom --exchange component of \sigpompom .
The solid line is the fit to the ``OR" points of a \regge --exchange
term, $(M_X^2)^{-0.32}$, added to this \pom --exchange term.
}
\label{fig:sigpp}
\end{figure}

We now ask how possible systematic uncertainties influence the
interpretation of the \sigpompom\ results shown in Fig.~\ref{fig:sigpp}.
A systematic uncertainty common to both sets of cross section results 
comes from the Monte--Carlo acceptance shown in 
Fig.~\ref{fig:mcacc}. 
We estimate that this uncertainty is smaller than 15\%. 
In the case of the ``AND" results, 
the statistical errors are much larger than this.

There is an additional systematic uncertainty that is very different
for the ``AND" and ``OR" results.
This arises from the possible non-universality of the \pom\ flux factor.
It is already known that \flux\ is not universal between $ep$ and $p\ap$ 
collisions because of the different effective \pom\ trajectory intercepts 
found in the two cases, attributable to different 
multi--\pom --exchange effects.  
Similarly, if multi-\pom -exchange is not identical in 
Reacts.~\ref{eq:pompom} and \ref{eq:dif}, there would be some
uncertainty as to whether the same flux factor should appear in
both Eqs.~\ref{eq:pompomdist} and \ref{eq:dsigdif}. 

We note, however, that this potential uncertainty does not exist for
the ``AND" results, because both final--state baryons have 
$|t| > 1.0$~GeV$^2$, where all the evidence\cite{es1,es2} 
points to an $s$--independent \pom\ trajectory; in that high--$|t|$ region, 
\flux\ appears to be insensitive to the damping which mainly leads to
an $s$--dependent effective \pom\ intercept\cite{es2} at $t = 0$.
Thus, the statistical errors on the ``AND" results in Fig.~\ref{fig:sigpp} 
dominate any uncertainties from this source.

The situation is very different for the ``OR" data sample because
the unseen final--state baryon has
low--$|t|$ and its flux factor in Eq.~\ref{eq:pompomdist} is sensitive to
the choice of effective $\epsilon$ value. 
We have thus recalculated the ``OR" cross sections in Fig.~\ref{fig:sigpp}
assuming a larger \pom\ trajectory intercept, $\epsilon = 0.10$, in \flux .
This is an extreme (unrealistic) 
case which assumes that there are no damping contributions from
multi--\pom --exchange in React.~\ref{eq:pompom} at our energy. 
We find that the ``OR" cross sections decrease from those shown in the figure;
for example, the lowest mass point decreases by 58\%, 
whereas the point at 11 GeV decreases by 41\%.
Thus, any increase in the effective $\epsilon$ used in calculating
\sigpompom\ from the ``OR" data
increases  the disagreement already observed in Fig.~\ref{fig:sigpp}.

The final systematic uncertainty comes from our model assumption that
there are no azimuthal angle correlations between final--state
$p$ and \ap\ in Eq.~\ref{eq:pompomdist}, other than the kinematic one
referred to in Sect.~\ref{sect:intro}.
However, the following two facts lead us to suspect that this may not be
true:  
(a) there are differences between the ``AND" and ``OR" \sigpompom\ 
results, and 
(b) the ``AND" and ``OR" data samples have very different \pom --\pom\ 
configurations.
The difference in \pom --\pom\ configurations 
is most easily visualized with reference to our Fig.~\ref{fig:dpevector}. 
In the upper figure (``AND"), it is seen that both \pom\ 
transverse momenta, \pt , are approximately in the same direction; 
hence $\Delta \pt \approx 0$.
In the lower figure (``OR"), one transverse momentum is near zero and the
other is near 1 GeV; hence the ``OR" data sample 
corresponds to $\Delta \pt \approx 1.0$~GeV.

Thus, the observed differences in \sigpompom\ for the ``AND" and ``OR" data 
in Fig.~\ref{fig:sigpp} at low mass suggest that \sigpompom\ depends 
on the \pom --\pom\ relative configuration and is larger at low
mass when the two \pom s move approximately in the same direction.
We also note that the systematic uncertainty due to our limited 
understanding of multi--\pom --exchange effects can not be responsible for
the effect now being discused, since it can only make the difference larger.
We return to the physics of the present discussion in the 
concluding Sect.~\ref{sect:discuss}.
 
\subsection{Test of factorization}
\label{sect:factor}
\indent

We now test the factorization relation, Eq.~\ref{eq:factorize}, between
\sigpompom , \sigpomtot\ and \sigpptot . 
If we multiply both sides of Eq.~\ref{eq:factorize} by $K^2$, we find:
\begin{equation}
K^2 \cdot \sigpompom (s') \, \, \, \, = \, \, \, \,
{  {[K \cdot \sigpomtot(s')]^2}  \over  { \sigma _{pp}^{total}(s') }        }.
\label{eq:Kfactorize}
\end{equation}
which relates precisely the measured quantities.
Thus, it is evident that tests of factorization using Eq.~\ref{eq:factorize} 
or~\ref{eq:Kfactorize} are equivalent and we may, with no loss
of precision, assume the value,
$K = 0.74$~GeV$^{-2}$ (see discussion in Sect.~\ref{sect:intro}),
and make the test using Eq.~\ref{eq:factorize}.

\begin{figure}
\begin{center}
\mbox{\epsfig{file=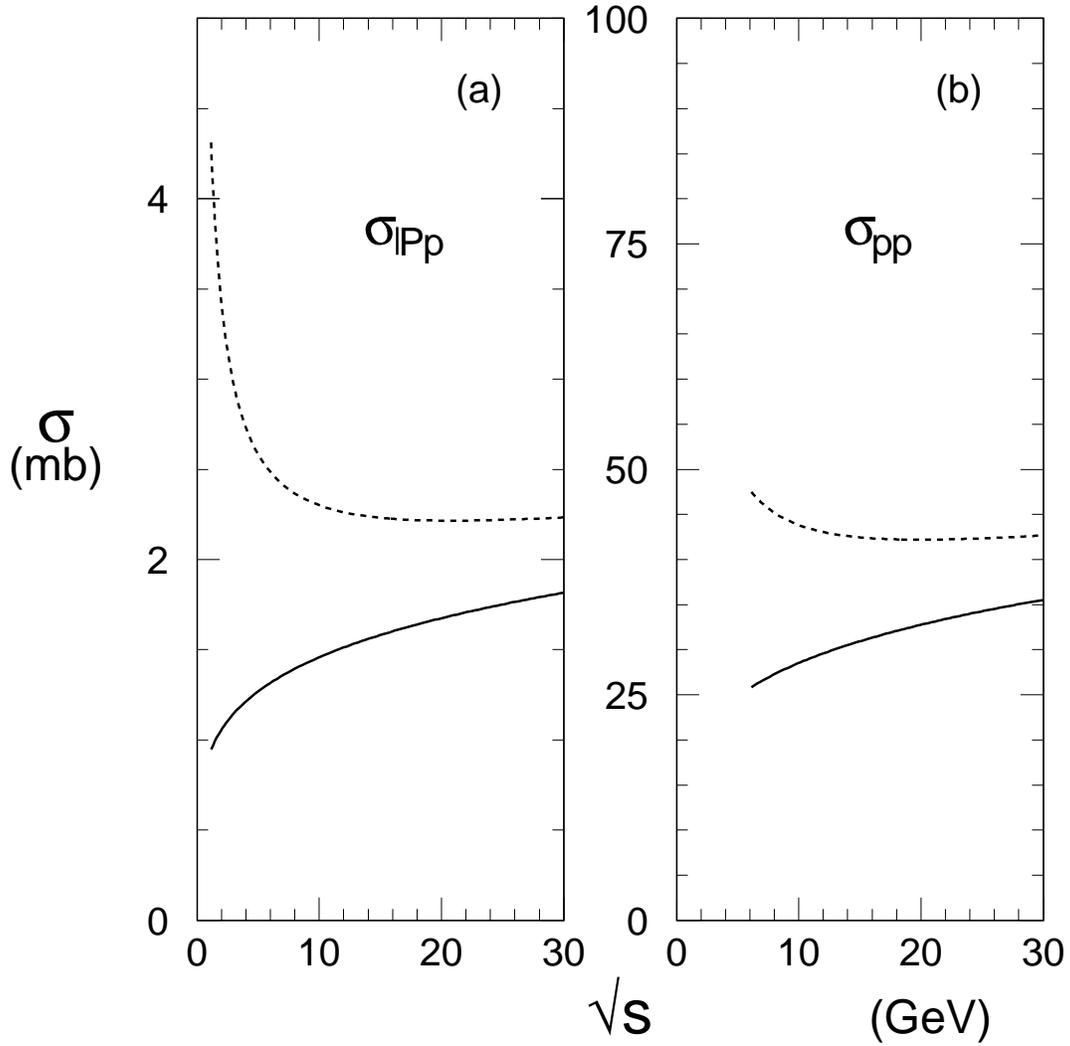,width=15cm}}
\end{center}
\caption[]{
(a) Dashed curve is \sigpomtot , the \pom -proton total cross section, from 
Ref.~\cite{ua8dif}, assuming $K = 0.74$~GeV$^{-2}$ (see text for explanation).
The solid curve is only the \pom -exchange component;
(b) same as (a), except for the proton-proton total cross 
section~\cite{cudell,covolan}.
}
\label{fig:totalsigs}
\end{figure}

In order to calculate the right-hand-side of Eq.~\ref{eq:factorize}
a function of $s'$, we use the following parametrizations
for the \pom --proton total cross section~\cite{ua8dif}:
\begin{equation}
\sigpomtot = {{0.72}\over{0.74}}
\cdot [(s')^{0.10} + 4.0 (s')^{-0.32}]\,\,\,\,\, {\rm mb}, 
\label{eq:spomprot}
\end{equation}
and for the proton-proton total cross section~\cite{cudell,covolan}:
\begin{equation}
\sigpptot =   18 \cdot s^{0.10} -27 \cdot s^{-0.50} 
+ 55 \cdot s^{-0.32}\,\,\,\,\,{\rm mb}.
\label{eq:sprotprot}
\end{equation}
These functions are shown as the dashed curves in 
Figs.~\ref{fig:totalsigs}(a,b).
Since Eq.~\ref{eq:factorize} is only valid for the \pom --exchange 
component of these functions, we show only the first terms in
Eqs.~\ref{eq:spomprot} and~\ref{eq:sprotprot} 
as the solid curves in the figures.

The dashed line in Fig.~\ref{fig:sigpp} shows the factorization
prediction for \sigpompom\
calculated using Eq.~\ref{eq:factorize} and the \pom\ terms in
Eqs.~\ref{eq:spomprot} and \ref{eq:sprotprot}:
\begin{equation}
\sigpompom (s') \, \, \, \, = \, \, \, \,
{{[\sigpomtot]^2} \over {\sigma _{pp}^{total}}} = 
{{(0.72/0.74)^2}\over{18}} \cdot (s')^{0.10}.
\label{eq:ffactorize}
\end{equation}
We see that there is increasingly better agreement between the prediction
and the measured \sigpompom\ points as the mass increases, as expected, since
the measured points contain both \pom\ exchange and \regge\ exchange.
The results are seen to be in reasonable agreement with the validity
of factorization for \pom -exchange in these reactions.

The solid curve in Fig.~\ref{fig:sigpp} is a fit to the ``OR" points with
$M_X > 10$~GeV of the sum of Eq.~\ref{eq:ffactorize} and a \regge --exchange
component as in Eq.~\ref{eq:spomprot}:
\begin{equation}
\sigpompom (s') \, \, \, \, = \, \, \, \,
{{(0.72/0.74)^2}\over{18}} \cdot  [(s')^{0.10} + R \cdot (s')^{-0.32}]. 
\label{eq:lasteq}
\end{equation}
We find a value, $R = 13.6 \pm 4.7$, with a $\chi^2/{\rm D.F.} = 1.3$.

\section{Conclusions and predictions}
\label{sect:discuss}
\indent

Table~\ref{tab:sigmas} demonstrates that there
is a new class of events, the so--called double--\pom --exchange (DPE) events
with characteristic rapidity--gaps in both arms, 
which appear when both \xp\ values are greater than 0.95.
The analysis which follows this observation shows that, remarkably, the 
Regge formalism describes all inclusive 
DPE and single--diffractive data, using the empirical
$s$-dependent effective \pom\ trajectory of Ref.~\cite{es2},
which is believed to be due to increasing multi-\pom --exchange effects with
energy~\cite{kpt}.
That Regge phenomenology works as well as it 
does, despite the complications of multi-\pom -exchange, 
should place constraints on a theory of such multi-\pom --exchange effects
yet to be developed.

We observe that the produced central systems in the DPE events display no 
anomalous multiplicity distributions or electromagnetic energy fraction of 
the total observed calorimeter energy. 
Our measurements agree with normal expectations. 

The main result of the work reported here is the \pom --\pom\ total
cross section, \sigpompom , shown in Fig.~\ref{fig:sigpp}.
At large mass, there is a statistically weak agreement with factorization 
predictions. 
However, for $M_X < 8$~GeV, both the ``AND" and ``OR" results exhibit large 
enhancements in \sigpompom , 
with the ``AND" result being about three-times larger than the ``OR" result.
There may be a dynamical reason for this, based on the fact that the
``AND" and ``OR" data sets have different \pom --\pom\ configurations.

These results are probably related to the WA102 Collaboration 
observations~\cite{WA102},
at the much lower  center--of--mass 
energy\footnote{This translates to \pom\ momentum fractions, $\xi$, which are 
22 times larger in WA102 than in UA8. Since this can lead to 
non--\pom --exchange contributions~\cite{ua8dif,albrowisr} as large as 50\% in 
WA102, we should not expect to observe exactly the same effects.}, 
$\sqrt{s} = 29$~GeV,
that the production of known quark-antiquark states and
glueball candidates depend differently on the difference in transverse 
momenta of the two \pom s in React.~\ref{eq:pompom}.
The primary WA102 effect is that the production of $q\bar{q}$ states appears 
to vanish as $\Delta \pt \ra 0.$ 
Close and Kirk~\cite{closekirk}
suggested that this effect may serve as a ``glueball filter".
On the basis of the WA102 results, 
Close and Schuler~\cite{cs} argue that the effective spin of the \pom\
can not be zero and that the \pom\ transforms as a non-conserved
vector current.

Taken at face value, the WA102 results~\cite{WA102} 
imply that our ``AND" data do not 
contain any $q\bar{q}$ states and, hence, that 
the enhancement for $2 < M_X < 8$~GeV 
in those data may be due to production of some glueball--like objects.
There would be a mix
of $s$-channel production (i.e. glueball alone) and production with other 
particles. The observation of resonant mass structure is precluded in the 
present experiment because of our poor mass resolution of $\approx 1.8$~GeV.

The next generation experiment of the type reported here 
should utilize a central detector capable of detailed studies 
of the produced central systems (including particle identification).
In order to be able to observe the azimuthal angle correlations of the
outgoing protons with properties of the central systems,
the Roman--pot systems (on both arms) should have as full an azimuthal
coverage as possible. 
 
\subsection{Predictions for Tevatron and LHC colliders}
\label{sect:predict}
\indent

The fact that the ``AND" and ``OR" data yield essentially the same 
\sigpompom\ at larger mass implies that our rapidity--gap procedures for
isolating React.~\ref{eq:pompom} are fundamentally sound and can used at
the much higher energy experiments at the Tevatron and LHC to look for
rare states with greatest sensitivity.

The study of the relatively pure gluonic collisions in React.~\ref{eq:pompom}
at higher energy colliders may yield surprising new physics.
In the UA8 papers on the occurrence and study of jet events in 
single--diffraction~\cite{ua8},
it was reported that, in about 30\% of the 2-jet events, the \pom\ appeared
to interact as a single hard gluon with the full momentum of
the \pom\ (the so--called ``Super--Hard" \pom ) .  This result suggests that,
in roughly 10\% of hard \pom --\pom\ interactions, there is 
effectively a gluon-gluon collision with the full $M_X$ of the central 
system. Thus, for example, at the LHC with $\rs = 14$~TeV, a rather pure sample
of central gluon-gluon collisions should occur with $M_X$ as large as
0.03(14) =  0.42~TeV (remember from Sect.~\ref{sect:intro} that we 
believe there is essentially pure \pom\ exchange at $\xi = 0.03$).

\begin{figure}
\begin{center}
\mbox{\epsfig{file=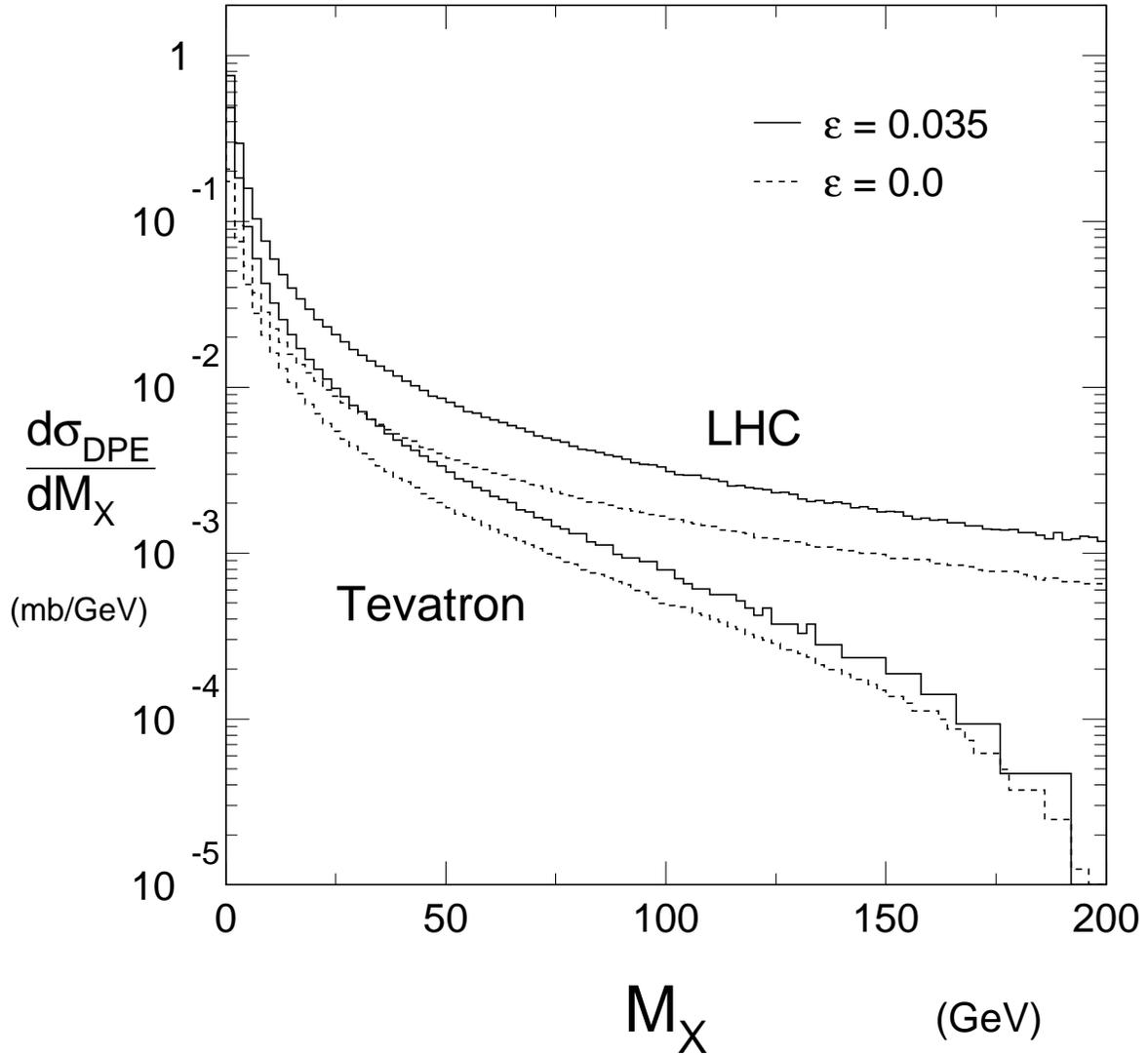,width=16cm}}
\end{center}
\caption[]{
Predicted differential cross section $d\sigma_{DPE} /dM_X$
(integrated over all $t$), assuming constant \sigpompom\ = 1 mb for the
Tevatron ($\rs = 2$~TeV) and the LHC ($\rs = 14$~TeV).
In each case, the solid (dashed) curves are for assumed effective \pom\ 
intercepts, $\alpha (0) = 1.035$ (1.00) respectively.
}
\label{fig:tevlhc}
\end{figure}

The phenomenology developed in our study of 
React.~\ref{eq:pompom} can be used to make cross section predictions at
the Tevatron ($\rs = 2$~TeV) and at the LHC ($\rs = 14$~TeV).
Figure~\ref{fig:tevlhc} and Table~\ref{tab:tevlhc}
show the results of Monte--Carlo calculations
of $d\sigma_{DPE} / dM_X$ (integrated over all $t$),
assuming that \sigpompom\ is $M_X$--independent and constant at 1~mb.
Since the fitted value~\cite{es2} 
of the effective \pom --Regge--trajectory intercept, 
$1 + \epsilon$ = 1.035, at \rs\ = 630 GeV, is also compatible with the 
available data at the Tevatron, we give the results for $\epsilon = 0.035$
at both Tevatron and LHC.
Schuler and Sj\"{o}strand~\cite{ss} suggest, in a model of hadronic diffractive
cross sections at the highest energies, that $\epsilon = 0$ 
is a reasonable approximation and we therefore also give results for this 
value. The observed peaking at small mass directly reflects the
$\xi$-dependence of the \pom\ flux factor in the proton.

To obtain cross section predictions for central Higgs 
production~\cite{dpehiggs} in React.~\ref{eq:pompom} from the
``Super--Hard" component of the \pom s ,
the calculations in Fig.~\ref{fig:tevlhc} can 
be multiplied by the 
calculated QCD cross section for the process (in units of mb)
based on the best available \pom\ structure function.

We note that the cross section predictions obtained in this way give
the total cross section for React.~\ref{eq:pompom}, where
there are no selection cuts on either final--state $p$ or \ap .
Clearly this yields the largest sensitivity for rare events.
In that case, if rapidity--gap vetos are used to suppress background
events, corrections must be made for the acceptance loss of central
system particles in the rapidity--gap regions.

\subsection{Predictions for forward spectrometers}
\label{sect:predict2}
\indent

For completeness, it may be useful to briefly summarize the possibilities for 
detection of DPE processes using existing or planned
forward multiparticle spectrometers.
There are two classes of such experiments, those traditionally called 
fixed--target experiments and those installed at 
storage-ring colliders\cite{pqs}.

Forward spectrometers installed at colliders can
observe DPE processes if there is an asymmetry between $\xi_1$ and $\xi_2$. 
The earliest example of such a measurement was Experiment R608 studying 
$pp$ interactions with $\sqrt{s} = 63$~GeV at the Cern 
Intersecting-Storage-Rings. 
Central production of $D(1285)$ was observed\cite{d1285} with almost pure 
helicity $\pm 1$, later explained by Close and Schuler~\cite{cs} as due to the
\pom\ behaving as a non-conserved vector current. 
In that process, the \pom\ appeared to dominate even though 
$\xi_1 - \xi_2 \approx 0.35$.
In the future, the higher energy
forward--spectrometer B--experiments, LHCb and B-TeV, 
will have good access to low-mass central systems with much smaller values
of $\xi$, as was pointed out in the LHCb Letter--of--Intent\cite{lhcb}.
 
\begin{figure}
\begin{center}
\mbox{\epsfig{file=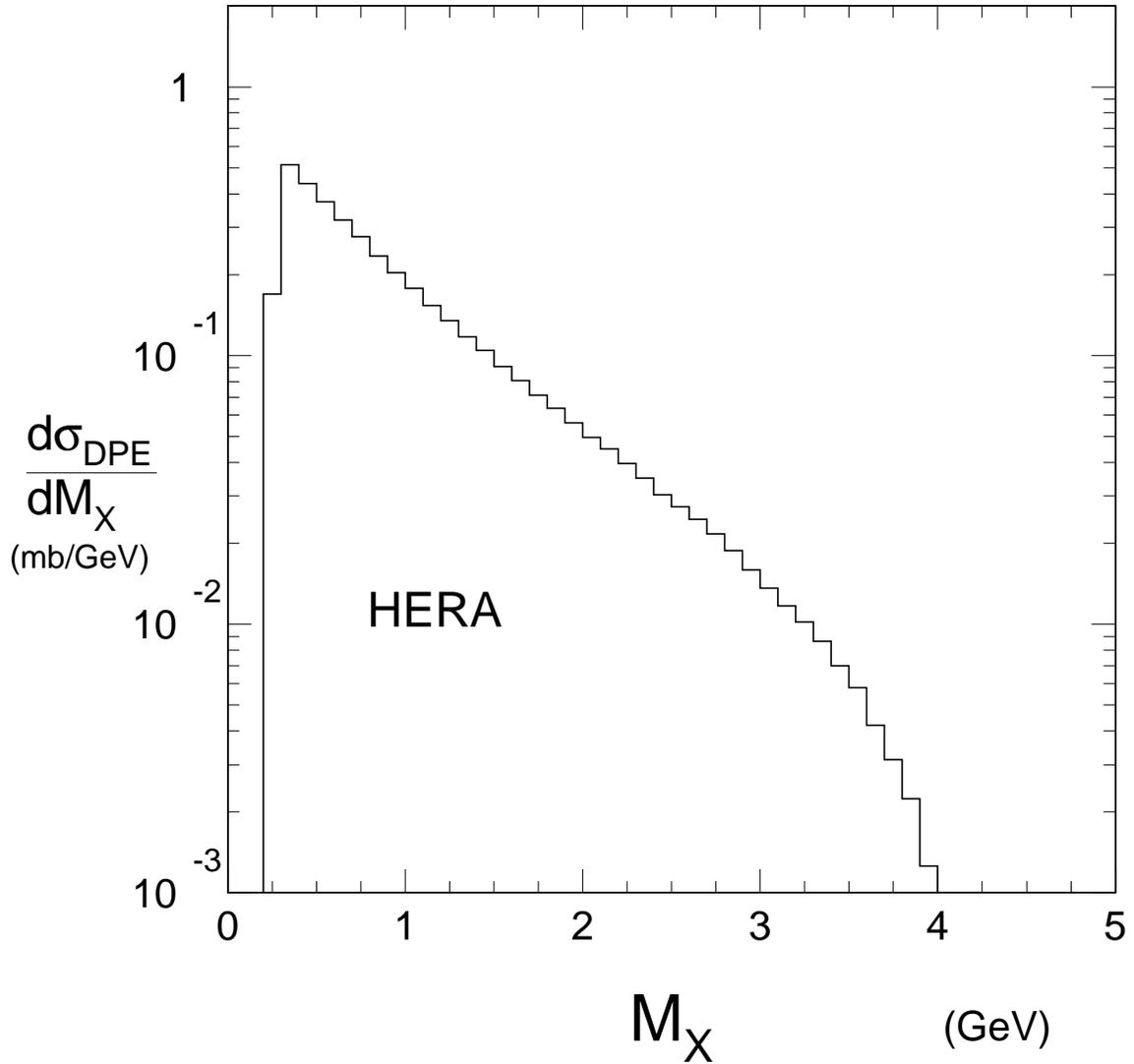,width=16cm}}
\end{center}
\caption[]{
Predicted differential cross section $d\sigma_{DPE}/dM_X$
(integrated over all $t$),
for the HERA-B experiment with $P_{beam} = 920$~GeV,
assuming fixed-target $pp$ interactions and constant \sigpompom\ = 1 mb.
}
\label{fig:hera}
\end{figure}

In fixed-target experiments, centrally--produced systems 
are boosted forward with the $\gamma$ of the center--of--mass, such that
$E_{Lab} = \gamma M_X$. Experiment WA102~\cite{WA102} was the first
experiment to carry out major DPE studies using this approach although,
as commented above, with a beam energy of 450 GeV, the \pom\ 
$\xi$--values were larger than desired.
We note that the existing experiment, HERA-B\cite{herab}, running 
at the HERA 920 GeV proton storage ring using wire targets,
could improve on the WA102 measurements.
For example, with $\sqrt{s} = 42$~GeV, production of a 2-GeV central system 
occurs with an average $\xi = 0.047$.  
Fig.~\ref{fig:hera} shows $d\sigma_{DPE}/dM_X$ 
for $pp$ interactions with a beam 
energy of 920 GeV, calculated as for Fig.~\ref{fig:tevlhc}. 
As indicated after Eq.~\ref{eq:traj}, the \pom\ trajectory intercept\cite{es2} 
used at this c.m. energy is: $1 + \epsilon = 1.087$.

\section*{Acknowledgments}

We appreciate many useful conversations with
Peter Landshoff, Mischa Ryskin and Alexei Kaidalov.
As this is the last of the UA8 publications, we wish to thank all those
who made it possible.
We are grateful to the UA2 collaboration and its original spokesman, Pierre 
Darriulat, who provided the essential cooperation and assistance.
We thank Sandy Donnachie for his strong support 
of the original UA8 proposal 
and the then-Director General, Herwig Schopper, for the support
of CERN.
We thank the electrical and mechanical shops at UCLA for essential
assistance in designing and building the UA8 triggering electronics
and wire chambers, and at Saclay for providing the UA8 trigger 
scintillators. The chambers were assembled and wired at CERN in the
laboratories of G. Muratori and M. Price
and we are indebted to them for their help.
We also thank Roberto Bonino and Gunnar Ingelman 
for their participation in the early stages of the experiment.



\newpage
 
\begin{table}
\centering
\begin{tabular}{|c||r|c|} 
\hline 
$p$ or \ap\ has $\xp > 0.95$  &Number  &Fraction with \\
Other has \xp\ in bin         &Events  &rapidity--gaps \\
\hline
0.70-0.75  &42   &$0.02 \pm 0.02$ \\
0.75-0.80  &113  &$0.04 \pm 0.02$ \\
0.80-0.85  &147  &$0.07 \pm 0.02$ \\
0.85-0.90  &163  &$0.03 \pm 0.02$ \\
0.90-0.95  &219  &$0.06 \pm 0.02$ \\
0.95-1.00  &314  &$0.48 \pm 0.04$ \\
\hline
\end{tabular}    
\caption[]{Numbers of events and their fractions 
which have rapidity--gaps, $2.3 < \eta < 4.1$ in both arms,
for different \xp\ selections of $p$ and \ap . 
1297 events have two reconstructed tracks
and at least 250~MeV of energy in the calorimeter system.
The table shows the 998 events, in which
either $p$ or \ap\ has $\xp > 0.95$,
while the other has \xp\ in the indicated bin.
In the remaining 299 events, for which both $p$ and \ap\ have 
$0.70 < \xp < 0.95$, only 0.7\% possess both rapidity--gaps.
}
\label{tab:sigmas}
\end{table}

\begin{table}
\centering
\begin{tabular}{|c||c|c||c|} 
\hline 
After              &Events       &percentage    &MC percentage \\
Cut                &remaining    &remaining     &remaining     \\
\hline
TOF veto           &188          &--            &--      \\
$\Phi_X$           &139          &$74 \pm 6$ \%  &72 \%   \\
\pt\               &126          &$67 \pm 6$ \%  &60 \%   \\
$\Sigma P_{long}$  &107          &$57 \pm 5$ \%  &50 \%   \\
\hline
\end{tabular}    
\caption[]{
Comparison of event losses for data and Monte--Carlo events (after TOF veto) 
as a function of the 3-momentum cuts in the event selection. 
}
\label{tab:mc}
\end{table}

\begin{table}
\centering
\begin{tabular}{|c||c|c||c|c|} 
\hline
$M_X$  &\multicolumn{4}{|c|}{$d\sigma_{DPE} / dM_X$    (mb/GeV)} \\ 
\hline
       &\multicolumn{2}{|c|}{Tevatron}  &\multicolumn{2}{|c|}{LHC} \\
(GeV)  &$\epsilon = 0.00$ &$\epsilon = 0.035$ 
       &$\epsilon = 0.00$ &$\epsilon = 0.035$      \\
\hline
  1    &1.25E-01        &2.69E-01       &2.07E-01       &7.55E-01 \\
  3    &5.44E-02        &1.00E-01       &9.42E-02       &2.96E-01 \\
  5    &3.01E-02        &5.17E-02       &5.39E-02       &1.58E-01 \\
 10    &1.31E-02        &2.11E-02       &2.53E-02       &6.76E-02 \\
 20    &5.33E-03        &7.65E-03       &1.15E-02       &2.75E-02 \\
 30    &3.02E-03        &4.14E-03       &7.17E-03       &1.62E-02 \\
 40    &1.96E-03        &2.57E-03       &5.08E-03       &1.13E-02 \\
 50    &1.40E-03        &1.78E-03       &3.86E-03       &8.34E-03 \\
 60    &1.01E-03        &1.25E-03       &3.11E-03       &6.45E-03 \\
 70    &7.76E-04        &9.42E-04       &2.56E-03       &5.23E-03 \\
 80    &6.08E-04        &7.15E-04       &2.17E-03       &4.32E-03 \\
 90    &4.78E-04        &5.54E-04       &1.85E-03       &3.75E-03 \\
100    &3.72E-04        &4.21E-04       &1.64E-03       &3.20E-03 \\
120    &2.36E-04        &2.48E-04       &1.28E-03       &2.46E-03 \\
140    &1.37E-04        &1.61E-04       &1.04E-03       &2.01E-03 \\
160    &7.45E-05        &9.35E-05       &8.87E-04       &1.60E-03 \\
180    &3.10E-05        &3.74E-05       &7.58E-04       &1.36E-03 \\
200    &1.12E-05        &4.21E-06       &6.55E-04       &1.18E-03 \\
\hline
\end{tabular}    
\caption[]{
Predictions for $d\sigma_{DPE} /dM_X$ (mb/GeV) at Tevatron and LHC 
assuming an $M_X$--independent \sigpompom\ = 1~mb,
for two values of effective \pom --trajectory intercept
(see Fig.~\ref{fig:tevlhc}.
}
\label{tab:tevlhc}
\end{table}


\clearpage

\clearpage

\clearpage

\clearpage

\clearpage

\clearpage

\clearpage

\clearpage
 
\clearpage
 
\clearpage
 
\clearpage
 
\clearpage
 
\clearpage
 
\clearpage
 
\clearpage
 
\clearpage

\clearpage
 
\clearpage
 
\clearpage

\clearpage

\clearpage
 
\clearpage
 
\clearpage
 
\clearpage
 
\clearpage
 
\clearpage
 
\clearpage
 
\end{document}